\documentclass[sigconf]{acmart}
\usepackage{soul}
\usepackage{graphicx}
\usepackage{subcaption}
\usepackage{enumitem}
\usepackage{fancyvrb, xcolor}
\usepackage{booktabs,multirow}
\usepackage{makecell}
\newcommand{\p}[1]{\ifx#1\empty\else$^{#1}$\fi} 

\newcommand{\RII}[1]{\textcolor{black}{#1}}
\newcommand{\RI}[1]{\textcolor{black}{#1}}
\newcommand{\RIII}[1]{\textcolor{black}{#1}}

\DefineVerbatimEnvironment{myprompt}{Verbatim}{
    rulecolor=\color{black},
    framerule=0.5pt,
    fontsize=\small,            
}

\AtBeginDocument{%
  }

\setcopyright{acmlicensed}
\copyrightyear{2026}
\acmYear{2026}
\setcopyright{cc}
\setcctype{by-nc-nd}
\acmConference[CHI '26]{Proceedings of the 2026 CHI Conference on Human Factors in Computing Systems}{April 13--17, 2026}{Barcelona, Spain}
\acmBooktitle{Proceedings of the 2026 CHI Conference on Human Factors in Computing Systems (CHI '26), April 13--17, 2026, Barcelona, Spain}
\acmPrice{}
\acmDOI{10.1145/3772318.3791777}
\acmISBN{979-8-4007-2278-3/2026/04}

\begin{document}
\newpage
\title{Does Personalized Nudging Wear Off? A Longitudinal Study of AI Self-Modeling for Behavioral Engagement}

\author{Qing He}
\authornote{Both authors contributed equally to this research.}
\orcid{0009-0005-1998-5772}
\affiliation{%
  \department{Weitzman School of Design}
  \institution{University of Pennsylvania}
  \city{Philadelphia}
  \state{Pennsylvania}
  \country{USA}
}
\affiliation{%
  \department{Key Laboratory of Pervasive Computing, Ministry of Education, Department of Computer Science and Technology}
  \institution{Tsinghua University}
  \city{Beijing}
  \country{China}
}
\email{qingh@upenn.edu}

\author{Zeyu Wang}
\orcid{0009-0007-5048-1665}
\authornotemark[1]
\affiliation{%
  \department{Department of Computer Science and Technology, Beijing National Research Center for Information Science and Technology (BNRist)}
  \institution{Tsinghua University}
  \city{Beijing}
  \country{China}}
\email{wang-zy23@mails.tsinghua.edu.cn}

\author{Yuzhou Du}
\orcid{0009-0000-7075-4252}
\affiliation{%
  \department{Grado Department of Industrial and Systems Engineering}
  \institution{Virginia Tech}
  \city{Blacksburg}
  \state{Virginia}
  \country{USA}}
\email{daviddu@vt.edu}

\author{Jiahuan Ding}
\orcid{0009-0002-6280-9713}
\affiliation{%
  \department{Joint School of Design and Innovation}
  \institution{Xi’an Jiaotong University}
  \city{Xi’an}
  \country{China}}
\email{dingjh@stu.xjtu.edu.cn}

\author{Yuanchun Shi}
\orcid{0000-0003-2273-6927}
\affiliation{%
  \department{Key Laboratory of Pervasive Computing, Ministry of Education, Department of Computer Science and Technology, BNRist}
  \institution{Tsinghua University}
  \city{Beijing}
  \country{China}
}
\affiliation{%
  \institution{Qinghai University}
  \city{Xining}
  \state{Qinghai}
  \country{China}
}
\email{shiyc@tsinghua.edu.cn}

\author{Yuntao Wang}
\authornote{Corresponding Author.}
\orcid{0000-0002-4249-8893}
\affiliation{%
  \department{Department of Computer Science and Technology}
  \institution{Tsinghua University}
  \city{Beijing}
  \country{China}
}
\affiliation{%
  \department{School of Computer Technology and Application}
  \institution{Qinghai University}
  \city{Xining}
  \state{Qinghai}
  \country{China}
}
\email{yuntaowang@tsinghua.edu.cn}

\renewcommand{\shortauthors}{He et al.}

\begin{abstract}
Sustaining the effectiveness of behavior change technologies remains a key challenge. AI self-modeling, which generates personalized portrayals of one’s ideal self, has shown promise for motivating behavior change, yet prior work largely examines short-term effects. We present one of the first longitudinal evaluations of AI self-modeling in fitness engagement through a two-stage empirical study. A 1-week, three-arm experiment (visual self-modeling (VSM), auditory self-modeling (ASM), Control; N=28) revealed that VSM drove initial performance gains, while ASM showed no significant effects. A subsequent 4-week study (VSM vs. Control; \RII{N=31}) demonstrated that VSM sustained higher performance levels but exhibited diminishing improvement rates after two weeks. Interviews uncovered a catalyst effect that fostered early motivation through clear, attainable goals, followed by habituation and internalization which stabilized performance. These findings highlight the temporal dynamics of personalized nudging and inform the design of behavior change technologies for long-term engagement.


\end{abstract}

\ccsdesc[500]{Human-centered computing~Empirical studies in HCI}
\ccsdesc[300]{Applied computing~Health informatics}
\ccsdesc[500]{Human-centered computing~HCI theory, concepts and models}

\keywords{AI Self-Modeling, Longitudinal Study, Nudging, Behavior Change Technologies}
\begin{teaserfigure}
  \includegraphics[width=\textwidth]{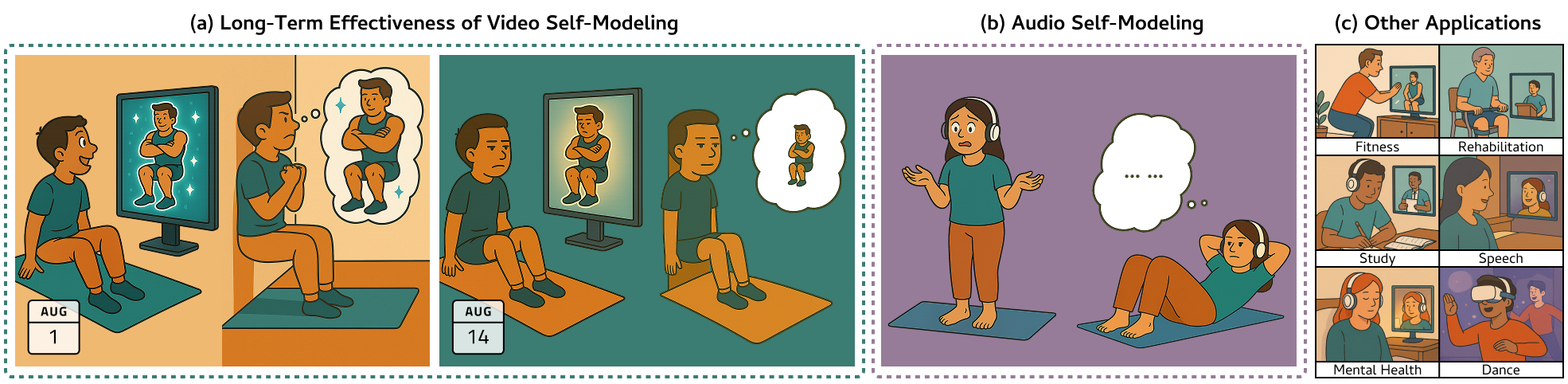}
  \caption{The nudging effect of AI self-modeling across time and modalities in fitness. (a) Change of nudging effect of video self-modeling(VSM): users are motivated by the AI self-model at first, then grow numb to the repeating VSM intervention, yet the motivated self-model has internalized as their goal.  (b) non-significant benefit of audio self-modeling (ASM) for fitness; (c) broader applications in other domains such as rehabilitation or learning.}
  \Description{Three scenario illustrations of AI self-modeling effects and applications. (a) Video self-modeling (VSM): participants are initially motivated by the AI avatar, later become desensitized to repetition, but gradually internalize the avatar as their goal. (b) Audio self-modeling (ASM): shows limited or non-significant fitness benefits compared to VSM. (c) Broader application domains: illustrates AI self-modeling applied to fitness, rehabilitation, study, speech, mental health, and dance.}
  \label{fig:teaser}
\end{teaserfigure}

\maketitle

\section{Introduction}


Long-term behavior change remains notoriously difficult. Even with the aid of personal informatics and behavior tracking tools, many users abandon their efforts after only a short time~\cite{gouveia_2015, shih_2015, lazar_2015}.  In response, researchers have looked toward \textbf{nudging}~\cite{thaler_2008}, a behavioral science approach that subtly guides individuals towards desired behaviors without restricting their choices~\cite{caraban201923ways, fogg2009behavior}. However, traditional technology-mediated nudges are relatively inflexible, with limited adaptation to individual needs and contexts~\cite{rapp2024open, chen2024investigating}. With recent advances in artificial intelligence (AI), new opportunities have emerged for performing more interactive, personalized, and context-aware interventions~\cite{rabbi2015mybehavior, jones2024designing, nakamura2023eat2pic, tian2021let}. A promising concept is \textbf{AI self-modeling}, which uses AI technologies to generate realistic, individualized simulations of a user's future self. By making long-term consequences vivid and personally relevant, it has the potential to sustain engagement, reinforce self-efficacy, and improve performance over time.

Early studies offer encouraging evidence across modalities. FakeForward, for instance, utilizes AI to swap faces onto a peer model with better performances in specific tasks~\cite{fakeforward}. Voice-based approaches, such as Emotional Self-Voice (ESV) and Mirai, use voice cloning technology to generate personalized and context-related audio responses~\cite{fang_2025_ESV, mirai}. These self-referential representations demonstrate the potential of AI self-modeling to boost motivation and performance. However, prior evaluations primarily focus on short-term effects, leaving open whether such benefits can be sustained in everyday contexts.

Psychological theories suggest that sustaining such effects over time is challenging. Habituation theory~\cite{rankin_2009, thompson_1966} suggests that repeated exposure reduces novelty and diminishes impact. In practice, fitness and health behavior change technologies (BCTs) face early attrition and engagement decay, driven by novelty decay, monitoring burden, lack of personalization, variety, and ongoing support~\cite{shin2019beyond, jeong2017smartwatch, gouveia_2015}. Affective–Reflective Theory also indicates that exercise decisions are often affect-driven in the short term, but sustained adherence requires reflective attitudes~\cite{brand_2016, brand_2018}. For AI self-modeling, which provides deeply self-referential, identity-linked feedback, engaging mechanisms distinct from generic prompts, its capacity to sustain nudging effects over time remains an important question.

To evaluate the pattern of long-term impact of AI self-modeling, our research focuses on fitness as a meaningful testbed whose objective performance and subjective experience are both measurable. Following previous works, we examine two representative AI self-modeling modalities: Video Self-Modeling (VSM)~\cite{fakeforward} and Audio Self-Modeling (ASM)~\cite{fang_2025_ESV, mirai}, providing an ideal self for participants during daily exercise. \RII{These implementations reflect common ways that self-modeling has been operationalized in practice, allowing us to situate AI-generated self-representations within everyday fitness routines.} To investigate their long-term effects, our study is guided by the following research questions:

\begin{enumerate}[label=\textbf{RQ\arabic*}]
  \item Does AI self-modeling sustain performance over time in everyday fitness practice?
  \item Does AI self-modeling sustain improvement rate over time in everyday fitness practice? 
  \item What design factors influence the long-term impact of AI self-modeling, and how can these insights inform future behavior change technologies?  
\end{enumerate}

We first conducted a one-week exploratory study (N=9,10,9) to evaluate both AI self-modeling techniques. The initial results showed that ASM failed to boost fitness performance, while VSM demonstrated effectiveness for one week. Based on the results, we expand the experiment period to four weeks for the VSM and the Control group \RII{(N=31)}. Our final results demonstrated that VSM could successfully sustain performance for four weeks, whereas it failed to sustain the improvement rate and showed a converging trend after two weeks. Further analysis of long-term subjective feedback revealed habituation to the continuous VSM intervention. Through semi-structured interviews with the participants, we discussed reasons for the results and provided in-depth design implications for future long-term nudging systems. 

We claim the following novel contributions of our work:

\begin{itemize}
    \item We present \RIII{the first 28-day long-term study of AI self-modeling with 31 participants, extending prior work~\cite{fakeforward,fang_2025_ESV} limited to short-term or one-off effects and revealing a stage-dependent trajectory, early acceleration followed by stabilization, which offers a more nuanced pattern than simple habituation decay.}
    \item Our studies offer design implications for \RI{long-term} behavior change technologies, reflecting how modality, personalization, and subjective experiences are crucial for designing interventions with lasting impact.
\end{itemize}

\section{Background and Related Works}
\RI{This section establishes the theoretical foundation for our work. We begin with the challenge of long-term behavior change and motivational decline, then introduce AI self-modeling and its theoretical potential to counteract such declines. Finally, we situate these mechanisms within the fitness context, which serves as the longitudinal testbed for our study.}

\RI{\subsection{Long-Term Behavior Change and Motivational Decline}} \label{sec:mot_decline}
Prior works in HCI have repeatedly highlighted the challenge of sustaining engagement with behavior change technologies~\cite{gouveia_2015, shih_2015, lazar_2015}. While mobile and wearable devices often spark strong initial interest, sustained use proves elusive: studies show that many users discontinue within weeks or months~\cite{yang_2020}. Research has shown that users often abandon devices once the novelty fades, when self-tracking becomes burdensome, or when the collected data no longer feels meaningful~\cite{clawson_2015}. Likewise, longitudinal studies emphasize that while BCTs such as goal-setting, feedback, or self-monitoring can spark short-term motivation, their benefits often taper off with repeated use~\cite{gouveia_2015}. 

To further explain this decline, researchers emphasize motivational deterioration as the key driver of disengagement~\cite{consolvo_2009, krzysztof_2014, paay_2015}. The Athlete Burnout model, for example, conceptualizes this process across three dimensions: emotional and physical exhaustion, reduced sense of accomplishment, and sport devaluation~\cite{raedeke1997athlete, raedeke2001development}, with the latter two being especially predictive of dropout~\cite{olsson2025multi}. Complementary perspectives point to mechanisms of attentional and affective decline: Habituation theory suggests that repeated exposure to the same stimulus gradually reduces its motivational salience~\cite{thompson_1966}, while Attention Restoration Theory argues that prolonged repetitive engagement depletes attentional resources and undermines motivation~\cite{kaplan_1995}. Affective–Reflective Theory emphasizes the relation between short-term affective impulses and longer-term reflective attitudes~\cite{brand_2016, brand_2018}. In digital and exercise contexts, these processes manifest as fading novelty effects and flattened engagement curves. 

These empirical findings and psychological theories reveal that declining engagement is not a superficial issue but a fundamental challenge of human motivation and attention. \RI{However, while these theories explain why motivation decays, they provide limited guidance on how technology might sustain motivation across repeated daily exposures. This motivates our examination of whether AI self-modeling, which is augmented with identity cues and novelty, can counteract such decline.}

\subsection{AI Self-Modeling as Personalized Nudging for Behavior Change}

\subsubsection{\RI{Existing Works on AI Self-Modeling}}
Nudging, a strategy for influencing behavior without restricting choice~\cite{thaler_2008, caraban201923ways, fogg2009behavior}, has become a prominent approach in HCI.
However, traditional technology-mediated nudges are often static and insensitive to individual differences~\cite{rapp2024open, chen2024investigating}. \RI{Generative AI now enables \emph{AI self-modeling}, which produces individualized portrayals of one’s future or ideal self to deliver identity-based nudges.}


Visual approaches synthesize personalized images or videos that depict users (or lookalikes) successfully executing target behaviors, such as aged future selves~\cite{hershfield2011increasing}, or increasing attainability perceptions through deepfaked peer demonstrations~\cite{fakeforward}. Audio-based approaches modified self-voices can alter emotions~\cite{costa2018regulating} and nudge people to achieve daily goals~\cite{kim2024my}, cloned voices can deliver motivational utterances~\cite{fang_2025_ESV}, and wearable “inner-voice” systems provide context-aware interventions~\cite{mirai}.
\RI{Such systems primarily operate through verbal persuasion and affective modulation, which may influence motivation differently from visual interventions.} 
Conversational future-self agents \RI{also} combine visual and audio modalities to reduce anxiety and strengthen identification with one’s future self~\cite{pataranutaporn2024future}. Across these modalities, the common principle is aligning intervention content with a user’s identity signals (“this is me”) to strengthen intention, self-efficacy, and adherence.

\RI{Despite these technological advances, empirical evaluations have been largely confined to short-term or one-off effects. Consequently, the longitudinal durability of these AI-driven interventions in everyday contexts remains underexplored.}


\RI{
\subsubsection{Theories and Principles Behind Self-modeling}
Self-modeling draws from multiple psychological frameworks, but its foundational mechanism is best described as feedforward, or ``learning from the future''~\cite{dowrick1999review, dowrick2012self}. Unlike traditional feedback, which corrects past errors, feedforward creates a memory of success that has not yet occurred. AI operationalizes this by synthesizing ``future-perfect'' representations, allowing users to sense their own potential. This feedforward mechanism influences long-term behavior through two complementary pathways: a capability pathway that shifts self-efficacy (how capable one feels), and an identity pathway that shifts the salience and relevance of valued future selves (who one feels they are becoming).}

\RI{\textbf{Self-Referential Modeling Improves Self-Efficacy.}
Within Social Cognitive Theory, self-efficacy is shaped by multiple sources, including vicarious experience, verbal persuasion, and affective states~\cite{bandura_1977}. Self-modeling can be understood as a form of \emph{self-referential modeling}, where curated representations of one’s own performance act as symbolic models that inform what one believes they can achieve~\cite{dowrick1999review, dowrick2012self, dowrick2012selffuture}.} 

\RI{\textbf{Identity-Based Motivation and Possible Selves.}
Beyond perceived capability, identity-based motivation~\cite{oyserman_2015} highlights that people are motivated to act in ways that are congruent with ``people like me,'' and that they are more likely to interpret effortful actions as meaningful when these actions fit a valued identity. Possible Selves Theory~\cite{markus_1986} further describes how mental representations of hoped-for and feared future selves provide a cognitive bridge between present choices and long-term outcomes. When future selves are vivid, plausible, and experientially close, they render long-term goals personally relevant and worthy of effort. }

\RI{Compared to traditional methods, AI self-modeling theoretically enhances these pathways by rendering future selves with greater vividness (strengthening self-efficacy) and dynamic coherence (preserving identity fit). At the same time, AI allows these self-representations to be flexibly updated to match a person’s current progress and goals, preserving a tight fit between the modeled capabilities and the identity that feels attainable and self-relevant. In this way, AI self-modeling has the potential to counteract habituation and sustain the vividness required for the identity pathway, which is under-explored in previous studies. } 


\subsection{Fitness and Health as Context for Behavior Change}
To investigate these longitudinal dynamics in a motivationally demanding environment, we situate our study in fitness and health. Fitness tasks inherently demand repeated and sustained effort, making them an ideal setting for studying long-term motivational dynamics. Prior work has leveraged fitness as a natural testbed: ~\citet{gouveia_2015} conducted a ten-month field study of an activity tracker and found that only 38\% of participants continued after the first week, with most dropping out within two weeks. 

Fitness and health also offer a methodological advantage: it enables systematic evaluation of both \textit{objective performance outcomes} (e.g., exercise completion, physical improvement) and \textit{subjective motivational experiences}. For example, ~\citet{choe_2014} showed how self-trackers reflect not only on numerical performance but also on motivational struggles in everyday practice. ~\citet{mazeas2022evaluating} used accelerometers, pedometers, and self-report questionnaires to collect subjective and objective data, measuring the outcome of gamification on physical activity.

\RI{Crucially, physical training naturally accommodates the two dominant forms of AI self-modeling: visual and auditory interventions. In exercise contexts, visual cues are intrinsic to motor learning and form calibration (e.g., video feedback~\cite{modinger2022video}), while auditory cues are fundamental for affective regulation and persistence (e.g., rhythm and self-talk~\cite{hardy2006speaking, hatzigeorgiadis2011self, bood_2013}). Consequently, fitness provides an ecologically valid environment to implement both VSM and ASM as natural, representative instances of the AI self-modeling paradigm. This allows us to examine the longitudinal efficacy of the AI self-modeling concept broadly, rather than being limited to a single modality.}


\section{AI Self-Modeling Nudging System Implementation}

\RI{Based on the discussion in Section~\ref{sec:mot_decline}, behavior change technologies (BCTs) suffer from habituation during long-term usage. The emergence of AI self-modeling offers greater vividness and flexibility to strengthen self-efficacy and identity-based motivation. Our work aims to investigate whether these unique properties of AI self-modeling have the potential to alleviate longitudinal motivation decay. Guided by this rationale, we implemented our system by reproducing and extending the pipelines from ~\cite{fakeforward, fang_2025_ESV}, deploying them for the first time in a long-term study.}
The system consists of two components: \textbf{Video Self-Modeling (VSM)} and \textbf{Audio Self-Modeling (ASM)}, \RI{which are the mainstream strategies in AI self-modeling.} In this section, we first describe our implementation for VSM and ASM, followed by the key modifications in our implementation compared to the original pipeline~\cite{fakeforward, fang_2025_ESV}. 

\begin{figure}[t]
  \centering
  \includegraphics[width=\columnwidth]{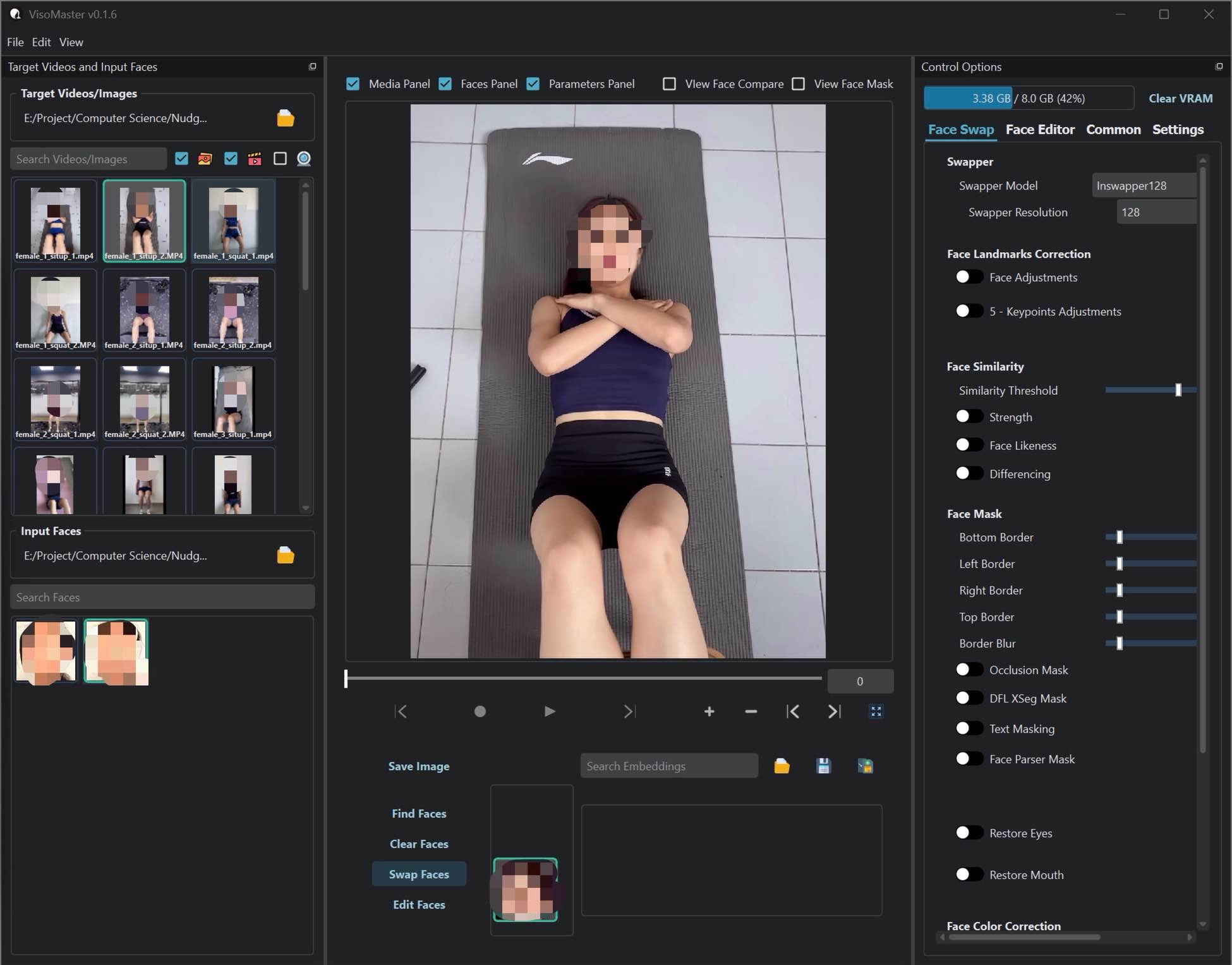}
  \vspace{2pt}
  \textbf{(a) Video Self-Modeling}
  \vspace{8pt}
  \includegraphics[width=\columnwidth]{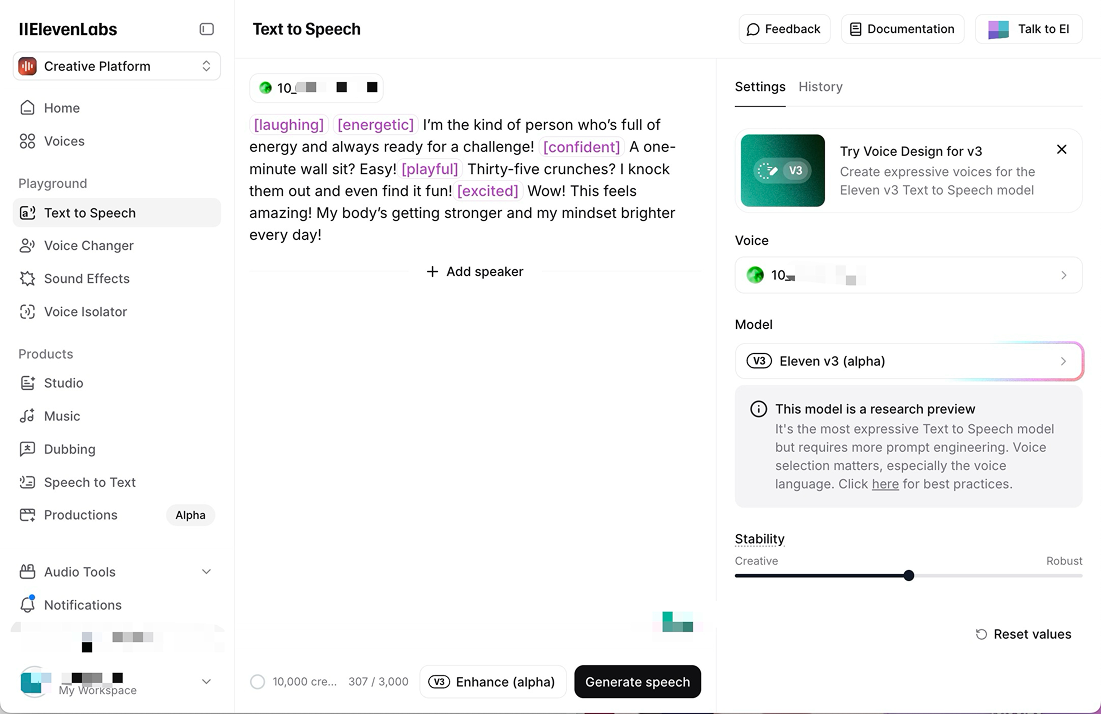}
  \vspace{2pt}
  \textbf{(b) Audio Self-Modeling}
  \caption{System implementation for AI self-modeling. Interfaces are adapted from existing platforms and included for illustration only.}
  \label{fig:system_implementation}
  \Description{Two images of the interfaces used for video and audio self-modeling. One shows the Visomaster interface, while the other shows the Elevenlabs web page.}
\end{figure}

\subsection{Video Self-Modeling Pipeline}
Our video self-modeling pipeline builds on the FakeForward framework~\cite{fakeforward}, which produces AI-edited future-self videos by swapping a participant's face onto a better-performing peer model. The key components of our implementation involve three stages:

\subsubsection{Peer Model Preparation.}
\label{sec:system_vsm_peermodel}
In this stage, we prepared high-quality peer model videos as face-swapping sources. Six participants (3 female, 3 male), each representing a different ideal body type, were recruited to demonstrate superior form and performance in target exercises. The recordings featured consistent lighting and camera angles to ensure seamless swapping, while varied clothing and backgrounds for each peer model enhanced resemblance and engagement.


\subsubsection{Headshot Capture and Personalization.} In this stage, the participant is guided to capture a headshot and select an ideal body type. Participants should ensure their face is clearly visible—avoid wearing glasses, having bangs or hair covering the face, and wearing makeup, to improve the quality of face-swapping.

\subsubsection{Face-swapping and Video Generation.} In the final stage, the participant's face is swapped onto the peer model's body using the open-source VisoMaster framework\footnote{\url{https://github.com/visomaster/VisoMaster}}, with built-in face detection and Inswapper128 as the swapper model (Figure~\ref{fig:system_implementation}a). During face-swapping, facial movements were synchronized with the peer model’s actions for a seamless effect, and videos were kept concise (30–40 seconds) to sustain attention.

\subsection{Audio Self-Modeling Pipeline}
The audio self-modeling pipeline adapts the ESV's approach~\cite{fang_2025_ESV}, which provides contextually aware interventions to support personalized goals. We adapt this method into the following steps:

\subsubsection{Voice Cloning.} Participants record a short audio clip by reading a script with colloquial happy and sad sentences in the local language (see Appendix~\ref{ap:scripts}).
Recordings were conducted in a quiet environment and then uploaded to ElevenLabs’ voice cloning service\footnote{\url{https://elevenlabs.io/voice-cloning}}. Participants were instructed to speak clearly at a moderate pace to ensure both intelligibility and emotional nuance were conveyed.


\subsubsection{Response Generation.} To adapt to participants' daily goals, which can be influenced by cognition and emotions, this stage generates motivational responses based on the participant's current context, which is captured through a daily Audio Generation Questionnaire (namely AGQ, see Appendix~\ref{ap:AGQ}), which contains the same fields as ESV~\cite{fang_2025_ESV} except for content choice. Due to the lack of API support at the time of the study, we did not re-implement its UI as the original pipeline did. Instead, we leveraged Azure GPT-4o's API~\footnote{\url{https://learn.microsoft.com/en-us/azure/ai-foundry/}} for generation and followed the prompt \RIII{structure} in ~\cite{fang_2025_ESV}.
\RIII{The prompt design (Appendix~\ref{ap:prompt}) ensured that generated messages reliably incorporated two validated psychological components: (1) positive self-talk from sports psychology~\cite{hardy2006speaking, hatzigeorgiadis2011self}, providing short directive cues known to enhance persistence (e.g., “you can hold this pace”); and (2) self-efficacy reinforcement grounded in social-cognitive theory~\cite{bandura_1977}, supporting participants’ belief in their ability to complete the exercise.}



\subsubsection{Audio Generation.} In this final stage, we leveraged ElevenLabs' V3 model to generate the motivational audio with emotions~\footnote{Eleven v3: \url{https://elevenlabs.io/text-to-speech}}. ElevenLabs' V3 can generate emotional audio controlled by open-vocabulary emotion description and special tokens. We generate four audio clips with emotional enhancement and choose the one that preserves the highest human-like quality and emotional expressiveness. 

\subsection{Adjustment to Original Pipeline}
To preserve comparability, we adhered closely to the original pipelines, matching their data preparation, prompt/script templates, and content scheduling, and introduced only two pragmatic substitutions to fit our study setting. For VSM, we replaced the old face-swapping backbone with a state-of-the-art model to improve identity fidelity and temporal coherence. For ASM, we adopted a speech-synthesis pipeline with stronger support for our local language to enhance naturalness and intelligibility. To measure the effectiveness of the self-modeling methods rather than the impact of specific content, we also introduced variation: VSM peer models prepared multiple outfits during filming, and ASM participants completed the AGQ daily to generate diverse audio clips.
These changes updated the tooling while keeping the experimental structure and stimulus intent intact.

\subsection{Ethical Concerns}
We treated identity editing and voice cloning as sensitive interventions and implemented safeguards addressing consent, privacy, fairness, transparency, and well-being in a unified protocol. All participants and peer models were provided explicit written consent in advance, specifying permissible uses (study-only, non-commercial), storage duration, etc. This research was reviewed and approved by the local Institutional Review Board.

\section{Study 1: Exploratory Validation of AI Self-Modeling}
We first conducted a one-week study to evaluate the effectiveness of video and audio AI self-modeling in our fitness context. \RI{The goal was to identify which implementation reliably produces measurable effects and could be extended to the subsequent long-term study.} In addition, we examined whether any nudging effects would diminish over time, as suggested by prior work on novelty decay and habituation.



\subsection{Study Design}
We designed a between-subjects experiment comparing the effects of daily exposure to AI-generated self-modeling media against a control condition. Participants were assigned to one of three groups based on their original physical condition: (1) video self-modeling (VSM), (2) audio self-modeling (ASM), or (3) Control (without nudging).

\begin{table}[t]
\caption{Demographic information of participants across three groups.}
\label{tab:demographics}
\Description{Demographic information of participants across the three groups (VSM, ASM, and Control). Each row reports gender distribution (female/male), group total, and mean BMI with standard deviation. All groups had balanced gender distributions, with 9–10 participants per group. Average BMI values were comparable across groups, ranging from 22.7 to 23.1, with no notable differences.}
\begin{tabular*}{\columnwidth}{@{\extracolsep{\fill}} l c c c @{}}
\toprule
\textbf{Group} & \textbf{Gender (F/M)} & \textbf{Total} & \textbf{BMI (M±SD)} \\
\midrule
VSM & 5/4 & 9 & 22.88±2.82 \\
ASM & 5/5 & 10 & 22.75±3.25 \\
Control & 6/3 & 9 & 23.11±2.91 \\
\midrule
Total & 16/12 & 28 & 22.91±2.90 \\
\bottomrule
\end{tabular*}
\end{table}

\subsubsection{Participants.} We recruited 30 participants (10 per group) with balanced gender distribution and similar BMI across groups to minimize potential confounding effects in physical performance. Two participants (one from the video group and one from the Control group) withdrew before the first intervention due to scheduling conflicts, leaving 28 participants for analysis (see Table \ref{tab:demographics} for demographics). Each participant received 20 USD per hour as compensation in accordance with local standards.

\subsubsection{Tasks.}\label{content:tasks}  
\begin{figure}[t]
    \centering
    \includegraphics[width=\columnwidth]{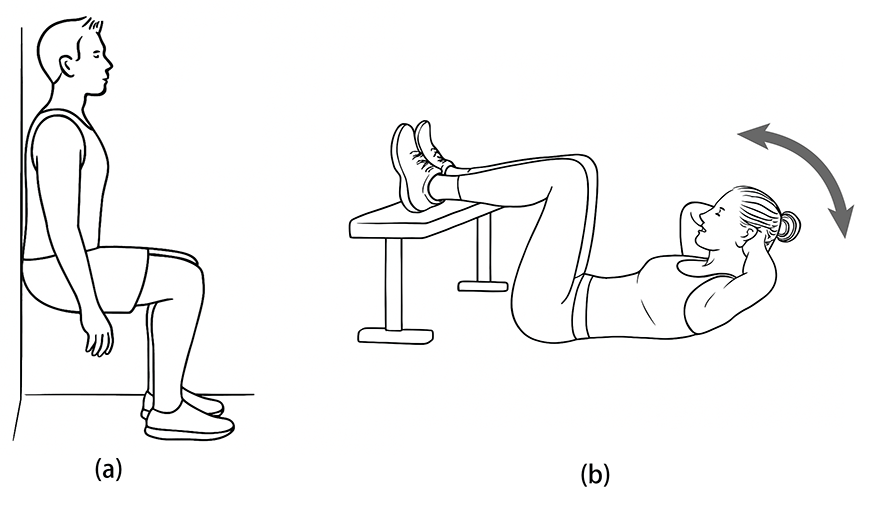} 
    \caption{Two selected exercising tasks: (a) wall-sit, (b) crunch.} 
    \label{fig:tasks}
    \Description{Two illustrations depicting the correct posture for the wall-sit and crunch tasks.}
\end{figure}

We selected two baseline tasks, \textit{wall-sit} and \textit{crunch}. \RII{Both exercises (1) offer clear, quantifiable metrics for daily longitudinal tracking, (2) were validated in prior self-modeling work~\cite{fakeforward}, and (3) target different muscle groups to avoid fatigue accumulation that could confound daily adherence: wall-sit for lower body and crunch for core/upper body.} Wall-sit (Figure~\ref{fig:tasks}a) is a static lower-body strength-endurance exercise in which participants slide down a wall until their hips and knees are flexed at a 90° angle and hold this position for as long as possible. In contrast, the crunch (Figure~\ref{fig:tasks}b) is a dynamic core and upper-body strength-endurance task. Participants lie supine with knees bent, cross their hands over their shoulders, and repeatedly raise their torso to a vertical position before lowering back down. Combining one static lower-body and one dynamic upper-body/core exercise distributes muscular load and minimizes fatigue, enabling sustained daily participation.

\subsubsection{Measurements.} \label{content:measurements} 
To measure the effectiveness of the nudging interventions, we utilized both objective and subjective measurements.

For objective measures, we recorded wall-sit duration and crunch repetitions. Participants performed both exercises once per day for 7 consecutive days, aiming to maximize repetitions for crunches and holding time for wall-sits. The results were recorded once they burned out or could not continue. To reduce individual variance, performance was normalized relative to each participant’s Day~1 value. Specifically, we analyzed percentage outcomes defined as $Pct_{i} = 100 \times \tfrac{x_{t}-x_{1}}{x_{1}}$, where $x_{t}$ is the performance value on day $t$ and $x_{1}$ is the Day~1 baseline. This transformation enabled us to model relative improvements over time, with $Pct$ serving as the dependent variable in our subsequent analyses.

For subjective measures, we administered three validated questionnaires adapted from prior works:
\begin{itemize}
  \item \textbf{Intrinsic Motivation Inventory} (IMI)~\cite{ryan_1982, ryan_1983} is used to measure intrinsic motivation and self-regulation~\cite{mcauley_1989}. We included the Interest/Enjoyment subscale (intrinsic motivation) and the Perceived Competence subscale (self-assessed task competence), rated on a 7-point Likert scale (see Appendix~\ref{ap:IMI}).
  \item \textbf{Exercise Self-Efficacy Scale} (ESES), originated from Bandura's self-efficiency theory~\cite{bandura_1977}, measures confidence in conducting activities under various conditions. ESES is an adjunct domain proposed in the Self-Efficacy Scale~\cite{bandura_2006} and has been validated by several studies~\cite{shin_2001, resnick_2000}. A 4-point Likert scale was used to rate the items (see Appendix~\ref{ap:ESES}).
  \item \textbf{Video/Audio Identification Questionnaire} (VAIQ), adapted from Player Identification Questionnaire ~\cite{vanlooy_2012} for avatars in online games. We applied Perceived Similarity, Embodied Presence, and Wishful Identification subscales to AI-generated video identification, while only Perceived Similarity for audio identification, rated on a 7-point Likert scale (see Appendix~\ref{ap:VAIQ}).
\end{itemize}

To minimize participant fatigue and inattentive responding, the full questionnaires were required on Day 1 (baseline) and Day 7 (post-intervention). On Days 2–6, participants completed shortened versions containing only essential items, while preserving psychometric validity.

\subsubsection{Procedure}
\begin{figure*}[t]
    \centering
    \includegraphics[width=\textwidth]{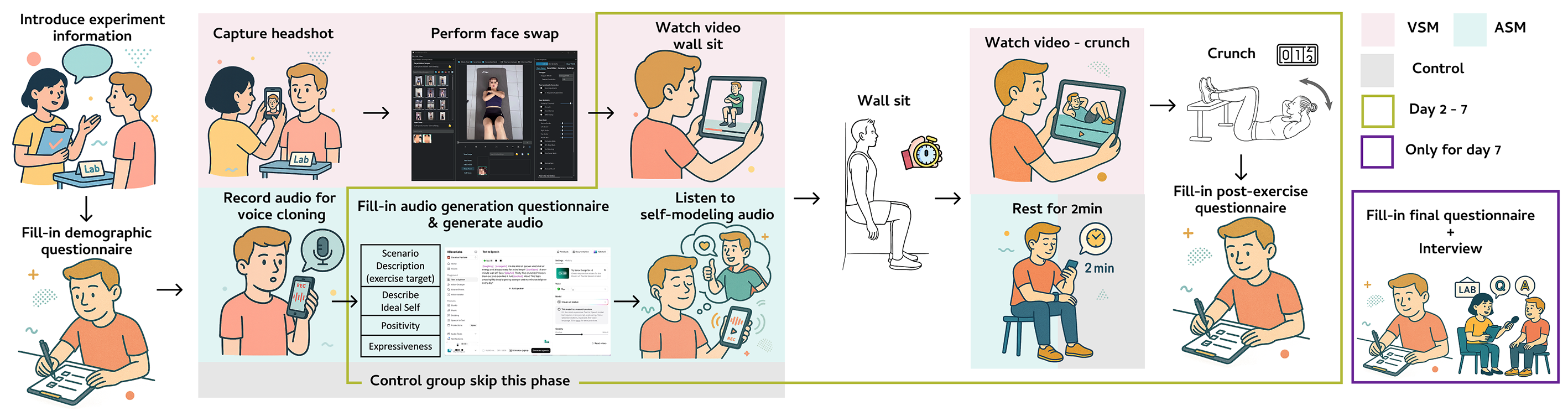} 
    \caption{Overview of the experimental procedure for three groups (VSM, ASM, and Control). The diagram illustrates the sequential tasks on Day 1 and subsequent sessions on Day 2-Day 7.} 
    \label{fig:study1_pipeline}
    \Description{Flowchart with fourteen scenarios connected by stages and flow links, showing the overall procedure of Study 1 for three groups (VSM, ASM, and Control). The start state is ‘introduce experiment information,’ and the end state is ‘fill in final questionnaire and interview.}
\end{figure*}

The study comprised three phases: Day 1 (lab session), Days 2–6 (remote sessions), and Day 7 (lab session). Figure~\ref{fig:study1_pipeline} shows an overview of the experimental procedures for task 1.

\textbf{Day 1 (Lab Session).} Upon arrival, participants provided informed consent, completed demographics (Appendix~\ref{ap:demographic}), and filled out the Physical Activity Readiness Questionnaire (PAR-Q) to ensure they could safely perform the exercises. Eligible participants then filled out the IMI and ESES. Group-specific preparations followed: the VSM group provided a headshot and chose a peer model for AI video generation; the ASM group recorded a script and submitted AGQs for voice cloning; the Control group skipped this step. To minimize expectancy effects, participants were not informed about the existence of other groups.

While waiting for the generation of AI contents, participants \RI{in all groups} were introduced to two target exercises, wall-sit (measured in duration) and leg-raise crunches (measured in repetitions). \RI{Then they viewed a brief example video demonstrating correct form and} received in-person posture guidance with practice trials. \RI{Afterward}, they were exposed to their assigned intervention:
\begin{itemize}
\item \textbf{VSM}: personalized AI-generated video (videos were watched before each exercise).
\item \textbf{ASM}: personalized AI-generated audio clip (audios were played once before all tasks).
\item \textbf{Control}: neutral pre-scripted verbal instructions.
\end{itemize}

Participants subsequently performed the wall-sit and crunch tests to exhaustion, with ~2 minutes rest between exercises. Performance was manually counted and recorded. A shortened IMI, ESES, and VAIQ (ASM, VSM only) were then required to fulfill after finishing the tasks.

\textbf{Days 2–6 (Remote Sessions).} Participants attended daily remote sessions via video-conference, scheduled at consistent times when possible. Before each session, the experimenter confirmed participants’ physical condition and absence of injuries. Then participants repeated the same exercise–intervention–exercise sequence as on Day 1. Objective performance was recorded remotely, and subjective measures were collected through online questionnaires. To ensure the variations in nudging content, experimenters randomly selected pre-generated peer model videos for VSM. While for ASM, participants are required to complete the AGQ 30 minutes before daily remote session for generating personalized audio clips.

\textbf{Day 7 (Lab Session).} Participants returned to the lab to repeat the Day 1 testing protocol, including full IMI, ESES, and VAIQ (ASM, VSM only) questionnaires. Finally, participants completed an open-ended interview to share subjective impressions of the intervention and study experience.

\subsection{Linear Mixed-Effects Analysis} \label{sec:exp1_analysis_methodology}
We employed Linear Mixed Effects models (LME) for data analysis. Specifically, we modeled participant $i$'s percentage performance $Pct$ (wall-sit or crunch performance relative to Day~1 baseline) on day $j$ as: 

\begin{equation}\label{eq:lme} 
\begin{aligned}
Pct_{ij} & = \beta_0 + \beta_1 t_{c,ij} + \beta_2 t_{c,ij}^2 + Group_{i} \times (\beta_3 + \beta_4 t_{c,ij} + \beta_5t_{c,ij}^2) \\
&+ u_{i} + \epsilon_{ij} 
\end{aligned}
\end{equation}

where $t_{c,ij}$ is mean-centered day, $Group_i\in\{0,1\}$ (0=Control, 1=VSM or ASM) denotes the intervention group, $u_i\sim\mathcal N(0,\sigma_u^2)$ is a participant random intercept, and $\epsilon_{ij}\sim\mathcal N(0,\sigma^2)$ is residual error. We estimated the fixed‐effect coefficients $\boldsymbol{\beta}=(\beta_0,\ldots,\beta_5)$, which quantify population-average effects on $Pct$.


\subsubsection{Test for RQ1 - sustained performance.}
The performance difference between the intervention group and the Control group can be formulated as: 

\begin{equation}
\begin{aligned}
\Delta(t)&= \mathbb{E}[Pct\mid Group{=}1]-\mathbb{E}[Pct\mid Group{=}0]\\
&= \beta_3+\beta_4 t_c+\beta_5 t_c^2
\end{aligned}
\end{equation}

where $t_c = t - c$, and $c$ is the mean study day. Choose an \textbf{early} reference time $t_E$ (the mean of an early window) and a \textbf{late} time $t_L$ (the mean of a late window). Define $t_{cE}=t_E-c$ and $t_{cL}=t_L-c$. We assess ``sustained performance'' with two linear-contrast tests from a single fit of Eq.~\ref{eq:lme}:

\begin{enumerate}
    \item \textbf{Late advantage (level at the end).} We test whether the intervention still outperforms Control at the late time:
   \begin{equation} \label{eq:analysis_rq1_lateadvantage}
       H_0:\ \Delta(t_L)\le0
   \quad\text{vs.}\quad
   H_1:\ \Delta(t_L)>0.
   \end{equation}
   \item \textbf{Sustainment (late $\ge$ early).} We test whether the gap at the end is no smaller than early on:
   \begin{equation} \label{eq:analysis_rq1_sustain}
   H_0:\ \Delta(t_L)-\Delta(t_E)<0
   \quad\text{vs.}\quad
   H_1:\ \Delta(t_L)-\Delta(t_E)\ge 0.
   \end{equation}

\end{enumerate}

We claim ``sustained performance'' if $\Delta(t_L)>0$ and $\Delta(t_L)-\Delta(t_E)\ge 0$ (CIs exclude $0$ in the one-sided direction).

\subsubsection{Test for RQ2 - sustained improvement rate.}
We operationalize ``sustained improvement rate'' as a non-decreasing change in slope difference between the intervention and Control groups over time. In contrast, we define ``\textbf{convergence}'' as a decreasing slope gap once the nudging effect has peaked. Drawing on theories like habituation, we hypothesize that long-term participants may become desensitized to the intervention, leading to convergence in the improvement trajectory.
In Eq.~\ref{eq:lme}, this is captured by the quadratic term $\beta_5$ ($Group \times t_c^2$), which governs how the slope differences evolve across days. 
We test convergence with the following equation: 
\begin{equation} \label{eq:analysis_rq2_beta5}
    H_0:\ \beta_5 \ge 0 \quad\text{vs.}\quad H_1:\ \beta_5 < 0
\end{equation}

\begin{table}[t]
\caption{Estimated fixed-effect coefficients ($\beta_0$–$\beta_5$) from Linear Mixed Effects models for VSM and ASM conditions.}
\Description{Estimated fixed-effect coefficients ($\beta_0$–$\beta_5$) from Linear Mixed Effects models for VSM and ASM conditions. Wall-sit task: Intercept terms are shared across groups with $\beta_0 = 13.58, \beta_1 = 13.06$, and $\beta_2 = 2.63$. For VSM, coefficients are $\beta_3 = 39.80, \beta_4 = 4.97, \beta_5=-2.61$. For ASM, coefficients are $\beta_3=-10.91, \beta_4=-7.10, \beta_5=-1.63$. Crunch task: Intercept terms are shared across groups with $\beta_0=20.45, \beta_1=6.75$, and $\beta_2=-0.08$. For VSM, coefficients are $\beta_3 = 30.75, \beta_4 = 9.79, \beta_5 = 0.26$. For ASM, coefficients are $\beta_3 = 7.12, \beta_4 = 3.69, \beta_5 = 0.64$.}
\centering
\resizebox{\columnwidth}{!}{
\begin{tabular}{lccccccc}
    \toprule
     & Group & $\beta_0$ & $\beta_1$ & $\beta_2$ & $\beta_3$ & $\beta_4$ & $\beta_5$ \\
    \midrule
    \multirow{2}{*}{\textbf{Wall-sit}} 
     & VSM & \multirow{2}{*}{$13.58$} & \multirow{2}{*}{$13.06$} & \multirow{2}{*}{$2.63$} & $39.80$ & $4.97$ & $-2.61$ \\
     & ASM & & & & $-10.91$ & $-7.10$ & $-1.63$ \\
    \midrule
    \multirow{2}{*}{\textbf{Crunch}} 
     & VSM & \multirow{2}{*}{$20.45$} & \multirow{2}{*}{$6.75$} & \multirow{2}{*}{$-0.08$} & $30.75$ & $9.79$ & $0.26$ \\
     & ASM & & & & $7.12$ & $3.69$ & $0.64$ \\
    \bottomrule
\end{tabular}}
\label{tab:exp1_analysis}
\end{table}

\begin{figure}[t]
    \centering
    \includegraphics[width=\columnwidth]{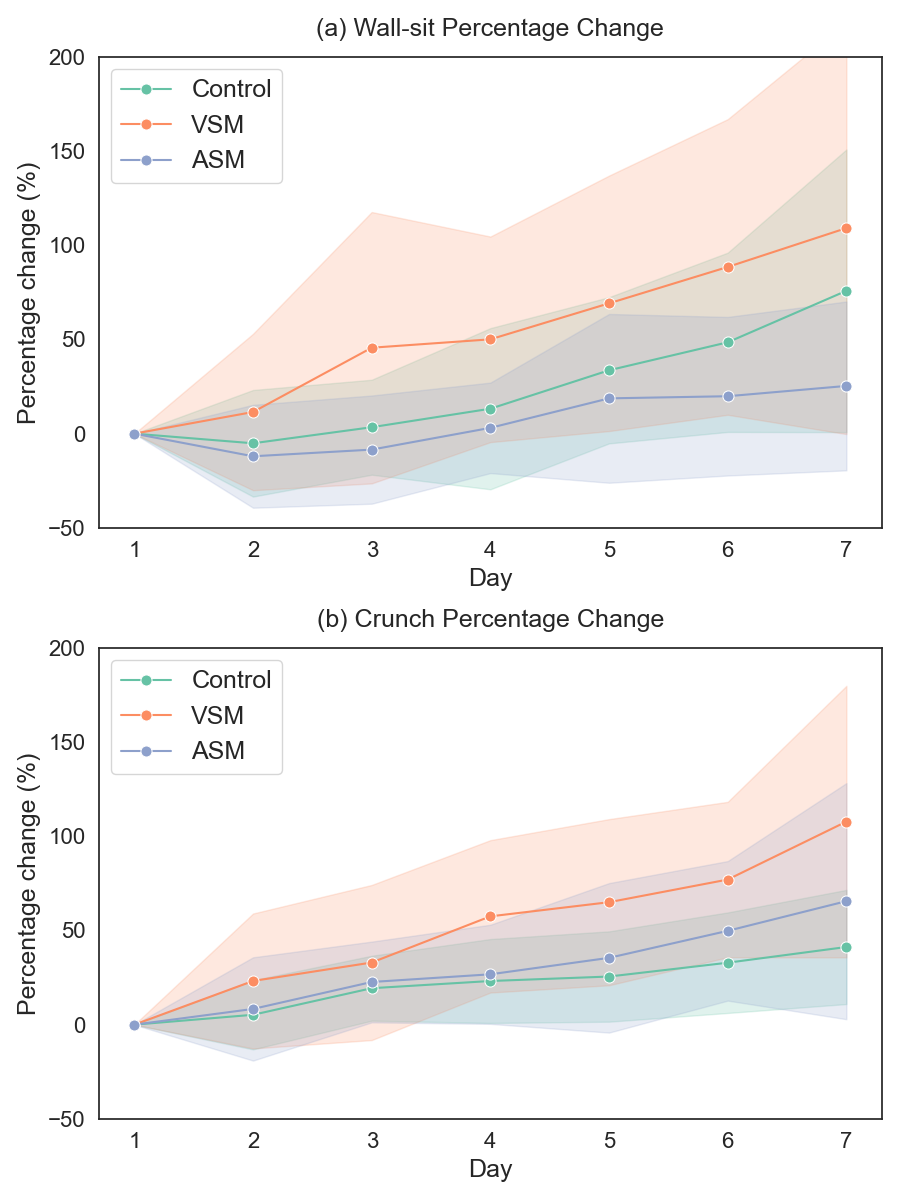} 
    \caption{Average performance change for VSM, ASM, and Control groups over 7 days.} 
    \label{fig:Study_1_performance}
    \Description{Two line graphs showing percentage performance change over seven days for VSM, ASM, and Control groups. (a) Wall-sit Percentage Change: VSM rises steeply, Control increases moderately, and ASM remains lower than Control. (b) Crunch Percentage Change: VSM improves most strongly, Control rises gradually, and ASM improves more than in wall-sit but remains below VSM.}
\end{figure}

\subsection{Results and Analysis}
The overall performance of all groups is illustrated in Figure~\ref{fig:Study_1_performance}, with model fitting results reported in Table~\ref{tab:exp1_analysis}. Within the 7-day intervention, the VSM showed faster early improvements, though the rate of gain tapered over time, providing only partial support for ``sustained performance''. In contrast, the ASM group demonstrated less consistent improvements. These patterns suggest that different self-modeling modalities may shape performance trajectories in distinct ways.  

\subsubsection{Understanding Control group's performance.}
The Control group displayed a steady upward trend throughout the week (Figure~\ref{fig:Study_1_performance}). The result of Mixed Linear Model regression confirmed significant positive linear effects of Time for the Control group on both wall-sit ($\beta_1=13.06, SE=4.67, p=.003^{**}$) and crunch ($\beta_1=6.75, SE=2.78, p=.008^{**}$). The findings indicate that participants in the Control group exhibit a gradual improvement in performance, reflecting natural practice and learning effects without any personalized intervention.

\subsubsection{RQ1: Does AI self-modeling sustain performance over time for both experiment groups?}
As shown in Figure~\ref{fig:Study_1_performance}, all groups improved on both wall-sit and crunch over the intervention period; the VSM group appeared to improve more rapidly, especially during the early days, while the ASM group exhibited less consistent gains. To evaluate the significance of VSM and ASM sustained performance gains, we tested both late advantages (end-of-week vs. Control) and sustainment (late vs. early advantage).  

For wall-sit, neither VSM nor ASM demonstrated significant advantages. VSM’s late advantage was positive but not reliable ($\Delta_L=31.18$, $SE=42.56$, $p=0.232$), and sustainment was also not significant ($\Delta_L-\Delta_E=29.81$, $SE=44.48$, $p=0.251$). ASM, by contrast, trended negatively, showing no late advantage ($\Delta_L=-47.25$, $SE=41.46$, $p=0.873$) and no sustainment ($\Delta_L-\Delta_E=-42.60$, $SE=43.35$, $p=0.837$). For crunch, VSM participants achieved a significant late advantage ($\Delta_L=62.46$, $SE=23.24$, $p=0.004^{**}$) and sustainment across the week ($\Delta_L-\Delta_E=58.71$, $SE=25.05$, $p=0.010^{**}$). ASM also showed a positive but weaker late advantage ($\Delta_L=23.93$, $SE=22.65$, $p=0.145$) and non-significant sustainment ($\Delta_L-\Delta_E=22.15$, $SE=24.41$, $p=0.182$) on performance. 

\RIII{The results indicate that AI self-modeling can support performance gains. At the same time, the weaker improvements observed for ASM may partly reflect the nature of the selected exercises, which rely heavily on posture and visual form. Because these tasks offer clearer visual than auditory information, different implementations of self-modeling may interact with the task context in distinct ways. We discuss this task-related limitation further in Section~\ref{sec:limitation}.}

\textbf{Key Findings:} VSM sustained performance gains for crunch but not for wall-sit, while ASM showed no reliable benefit compared to Control. 

\subsubsection{RQ2: Does AI self-modeling sustain improvement rate over time for the VSM group?}
To test whether VSM affected improvement rate (slope) rather than only level, we examined the quadratic interaction term ($\beta_5$) for $Group \times t_c^2$. A significant negative $\beta_5$ indicates convergence, reflecting an unsustained improvement rate relative to the Control as the nudging effect diminishes.


For crunch, $\beta_5=0.26$ ($SE=0.97$, $p=0.606$), providing no evidence of convergence under Eq.~\ref{eq:analysis_rq2_beta5}, and likewise non-significant when testing the alternative ($\beta_5>0$). In contrast, wall-sit ($\beta_5=-2.61$, $SE=1.32$, $p=0.023^*$) showed significant convergence. These results indicate that VSM did not sustain an accelerating improvement rate; instead, performance gains decelerated and converged toward Control levels, particularly in wall-sit. This pattern suggests that while self-modeling can boost early improvement, its rate effects diminish over time, reinforcing the importance of testing long-term trajectories in Study~2.  

\textbf{Key Findings:} VSM intervention demonstrates a decreased improvement rate for wall-sit during one week of exercise, but not for crunch.

\subsection{Conclusion and Implication for Study~2}
Overall, the results provide partial evidence addressing our research questions. VSM improved crunch performance and sustained outcomes, but wall-sit effects were inconsistent, showing convergence by Day 6. In contrast, ASM provided no reliable benefits in this context. 


\subsubsection{Interpretation of ASM effects.}
Compared to VSM, the method of audio intervention was far less effective in this scenario. Several explanations emerged from both the performance data and follow-up interviews. \textbf{First}, many participants reported that the generated audios sounded funny or even distracting, which made it difficult for them to take the feedback seriously or recall it during exercise. \textbf{Second}, participants often perceived the audio feedback as a simple emphasis on their goals rather than as a meaningful representation of their progress. As a result, the audio cues provided only limited encouragement. \textbf{Third}, unlike video feedback, which directly visualizes one’s body movements and progress, audio is relatively abstract and less embodied, reducing the potential to enhance their self-efficacy and engagement in training contexts. Taken together, these factors may explain why ASM produced little improvement beyond natural practice effects.

\subsubsection{Implication for Study~2}
For \textbf{RQ1}, we can partially conclude that VSM nudging can sustain performance for crunches over one week. For \textbf{RQ2}, VSM nudging did not sustain an improvement rate and instead demonstrated signs of convergence on wall-sit. However, as only one task reached statistical significance, the evidence is not strong. Moreover, one week of exposure might not be sufficient for a long-term study~\cite{del2024comprehensive}. Therefore, we recruited more participants and expanded the experiment cycle to 4 weeks in Study~2. Additionally, through ASM participants' feedback and performance, we discarded the ASM group in Study~2.

\section{Study~2: Longitudinal Effects of Video Self-Modeling}
Building on the findings from Study~1, we designed a one-month longitudinal study to examine the durability of AI self-modeling and focused exclusively on VSM. Extending the duration allowed us to capture performance  and improvement rates trajectories over time, and to test whether the nudging effect of AI self-modeling can persist in everyday fitness practice.



\subsection{Study Design}
This subsection details the participant recruitment and study procedure.

\begin{table}[t]
\centering
\caption{\RII{Summary of participants in each cohort by experimental condition.}}
\Description{Summary of participants in each cohort by experimental condition. Cohort 1 has 9 participants in the Control condition and 9 participants in the VSM condition, totaling 18. Cohort 2 has 6 participants in the Control condition and 7 participants in the VSM condition, totaling 13.}
\label{tab:cohort_summary}
\begin{tabular*}{\linewidth}{@{\extracolsep{\fill}} l c c c @{}}
\toprule
\textbf{Cohort} & \textbf{Control} & \textbf{VSM} & \textbf{Total}\\
\midrule
Cohort 1 & 9 & 9 & 18\\
Cohort 2 & 6 & 7 & 13\\
\bottomrule
\end{tabular*}
\end{table}

\subsubsection{Participants.}

\begin{figure}
    \centering
    \includegraphics[width=\columnwidth]{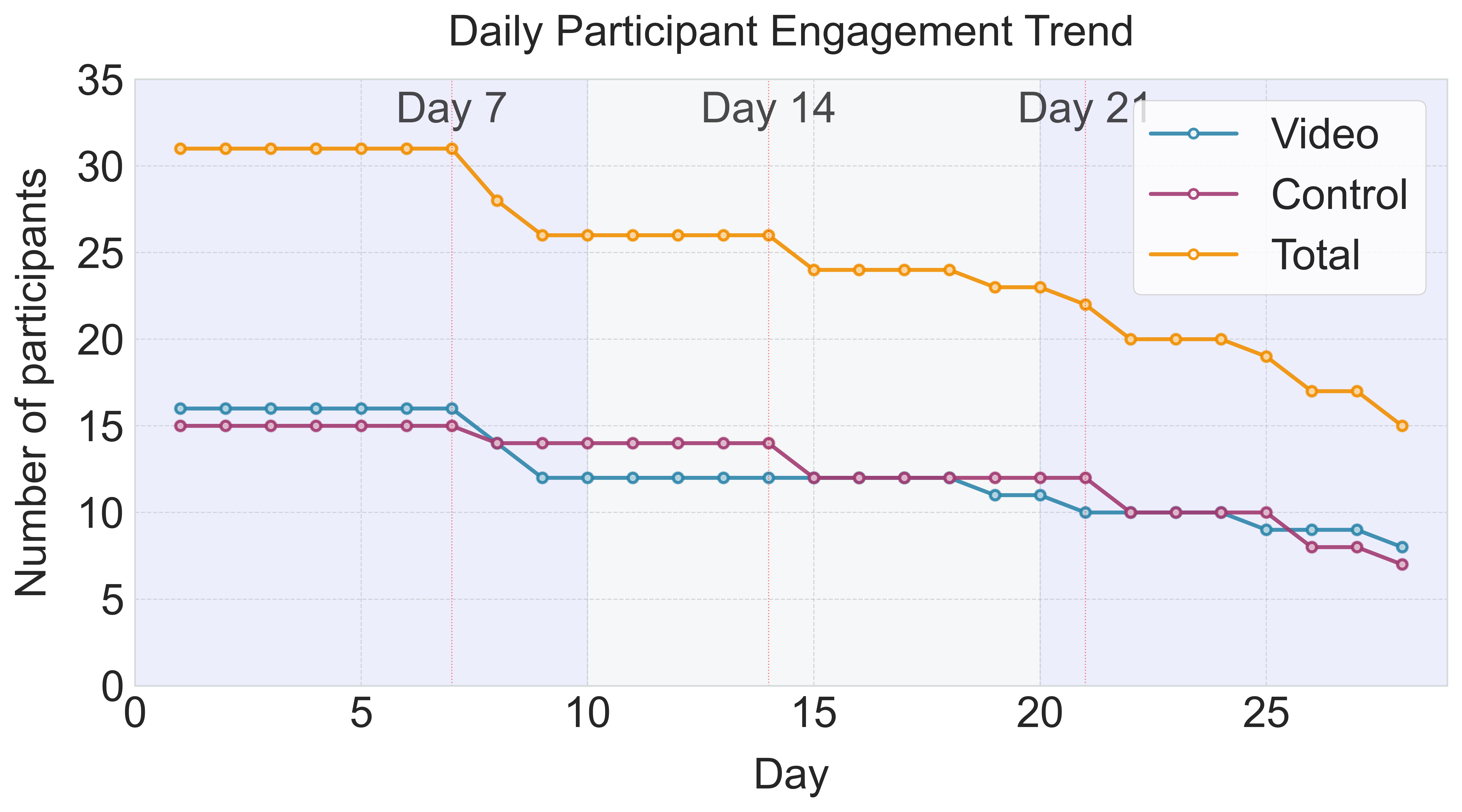} 
    \caption{\RII{Daily active participants during the 28-day intervention.}} 
    \label{fig:daily_participants}
    \Description{Line graph showing the number of daily active participants over the 28-day intervention. Total participants start at 32 and decline steadily to about 15 by Day 28. The Video group starts at 17 and drops to around 8, while the Control group starts at 15 and drops to around 7.}
\end{figure}

\RII{This study was composed of 31 participants in total. Among these 18 participants were from Study~1, of whom we referred to as Cohort~1; we also included 13 new participants, whom we referred to as Cohort~2 (see Table~\ref{tab:cohort_summary}). Study~2 aims to track performance over 4 weeks; thus Cohort~1 was expected to complete three additional weeks, and Cohort~2 was for four weeks, both following the same procedure. Each participant was compensated 20 USD per week.}
\RII{Figure~\ref{fig:daily_participants} shows the engagement gradually declined from 31} to 15 participants by the end, a typical attrition pattern in long-term interventions.


To ensure comparability, all core elements, including \textit{peer models} (Section~\ref{sec:system_vsm_peermodel}), \textit{tasks} (Section~~\ref{content:tasks}), and \textit{outcome measures} (Section~~\ref{content:measurements}), were kept identical to Study~1. The only difference was a short gap for Cohort~1 between studies, whereas Cohort~2 began without interruption. As shown in gap validation analysis (see Section\ref{content:gap_validation}), this gap did not introduce systematic bias, allowing the two cohorts to be pooled for further analysis.

\subsubsection{Procedure.} 
Informed by participants' feedback from Study~1, we conducted Study~2 entirely through remote sessions. Participants followed the structured rules to ensure consistency and reliability: 

\begin{itemize}
  \item \textbf{Voluntary participation:} Unlike Study~1, participants could withdraw at any point during the one-month period. 
  \item \textbf{Weekly unit:} Compensation was calculated by full weeks only; partial weeks were not compensated. 
  \item \textbf{Autonomy:} Exercises were completed independently at home without direct monitoring from researchers. 
  \item \textbf{Recording protocol:} Participants submitted daily video recordings in a standardized format: began with a phone screen showing the current local time, followed by continuous footage of the session. For VSM participants, the video-watching phase was also included.
\end{itemize}

\textbf{Onboarding session.}
For cohort~2, we conducted an online onboarding session to introduce the study protocol, including the nudging intervention (VSM only), tasks, and recording requirements. Participants also completed the baseline questionnaires (IMI, ESES, and demographics) and a remote Physical Activity Readiness questionnaire to confirm eligibility. 

VSM participants selected a preferred peer model and submitted a high-resolution headshot, which was used to generate four personalized videos (two wall-sits, two crunches) with varied contextual settings. The Control group relied solely on instructions from researchers without AI-generated content. 


\textbf{Daily routine.}  
Each day, participants completed the following procedures independently at home: \textbf{First}, perform \textit{wall-sits} to exhaustion (preceded by watching a wall-sit video for VSM, or following researcher instructions for the Control); \textbf{Second}, rest for two minutes; again perform \textit{crunches} to exhaustion following the same intervention for each group; \textbf{Finally}, upload the recording to the Cloud Drive and complete a daily questionnaire on objective performance and subjective motivation.

\textbf{Post-study session.}  
At the end of the period, participants completed the full IMI, ESES, and VAIQ (where applicable) questionnaires again, followed by an open-ended survey about their experiences and perceptions of the intervention.  

\subsection{\RII{Data Integrity and Attrition Analysis}}
Before the main analysis, we conducted \RII{three} checks to validate the effectiveness of our data. \RII{First, we examined whether merging the two cohorts (continuing vs. new participants) was statistically justified. Second, we rigorously assessed attrition patterns and potential survivor bias to ensure that dropout did not confound the comparison between intervention conditions. Third}, we assessed the reliability of the subjective scales on the pooled data, confirming internal consistency across items. 

\subsubsection{Gap Validation for Cohorts.}
\label{content:gap_validation}
To examine whether the gap between Study~1 and Study~2 introduced systematic bias in Cohort~1, we compared early improvement slopes between Cohort~1 and Cohort~2 during the initial phase of Study~2 by including Cohort in Equation~\ref{eq:lme} as a variable:
\begin{equation} \label{eq:gap_valid}
\begin{aligned}
Pct_{ij}=\gamma_0 
&+ \gamma_1 t_{c,ij} 
+ \gamma_2 t_{c,ij}^2 
+ Cohort_{i} \times (\gamma_3 
+ \gamma_4 t_{c,ij} 
+ \gamma_5 t_{c,ij}^2) \\
&+ Group_{i} \times (\gamma_6   
+ \gamma_7 t_{c,ij}
+ \gamma_8 t_{c,ij}^2)  
+ u_{i} + \epsilon_{ij} 
\end{aligned}
\end{equation}

where $Cohort_i \in [0,1]$ denotes cohort~1 or not. We fitted LME with performance data from Days~1-14, centering time at Day~7 so that the $Cohort \times t_c$ interaction represented the instantaneous slope difference at Day~7. For crunches, the Day~7 slope difference between cohorts was small and non-significant ($\gamma_4= -0.70, SE=0.53, p=0.182$); for wall-sit, it was also not significant ($\gamma_4= -1.30, SE=1.88, p=0.491$). Thus, there was no evidence of systematic cohort bias, justifying the pooling of Cohort 1 (continued) with Cohort 2 (new recruited) for the subsequent longitudinal analyses.

\RII{
\subsubsection{Attrition and Survivor Bias Analysis.}
We first examined whether attrition differed across cohorts or intervention groups, as differential dropout could bias longitudinal estimates. We compared (1) dropout proportions across $Cohort\times Group$ combinations using a Fisher exact test, and (2) attrition timing across the 28-day period using a Kaplan–Meier log-rank test. Both analyses showed no significant differences (Fisher exact $p = 1.00$; log-rank $p = 0.820$), indicating that neither cohort nor intervention condition exhibited distinct survival patterns.} 

\RII{To further rule out survivor bias (i.e., whether only highly motivated participants remained), we conducted a split analysis comparing the baseline characteristics (intrinsic motivation and Week-1 average performance) of \textit{Completers} (who finished the 4-week study) versus \textit{Dropouts} separately for each condition. For the VSM group, independent t-tests revealed no significant differences between Completers and Dropouts across all baseline metrics ($p > 0.050$), indicating that attrition was non-systematic and the intervention was effective across a broad range of users.}

\RII{In the Control group, while baseline physical performance was similar ($p > 0.050$), Completers had significantly higher IMI-Interest /Enjoyment ($p < 0.001^{**}$) and Competence ($p = 0.042^{*}$) than Dropouts. This result implies a conservative comparison for our main findings: the retained Control participants represented a subset of highly motivated individuals. Consequently, any performance advantage observed in the VSM condition would be achieved against a ``highly motivated'' Control baseline, thereby reinforcing the robustness of the intervention's effect.}

\RII{Detailed statistical tables and survival curves are provided in Appendix~\ref{ap:attrition}.
}

\subsubsection{Subjective Data Consistency Validation.}
We evaluated the reliability of subjective scores for IMI, ESES, and VAIQ using Cronbach’s $\alpha$, a standard index of internal consistency within a scale~\cite{cronbach_1951}. Values above 0.70 are generally considered acceptable, with higher values indicating stronger reliability. IMI reached excellent internal consistency for both Interest ($\alpha=0.83$) and Competence ($\alpha=0.86$) subscales; ESES also showed strong reliability ($\alpha=0.92$). For VAIQ, all dimensions demonstrated extremely high internal consistency with $\alpha$ values approaching 1.0.


\begin{figure*}[t]
    \centering
    \includegraphics[width=0.8\textwidth]{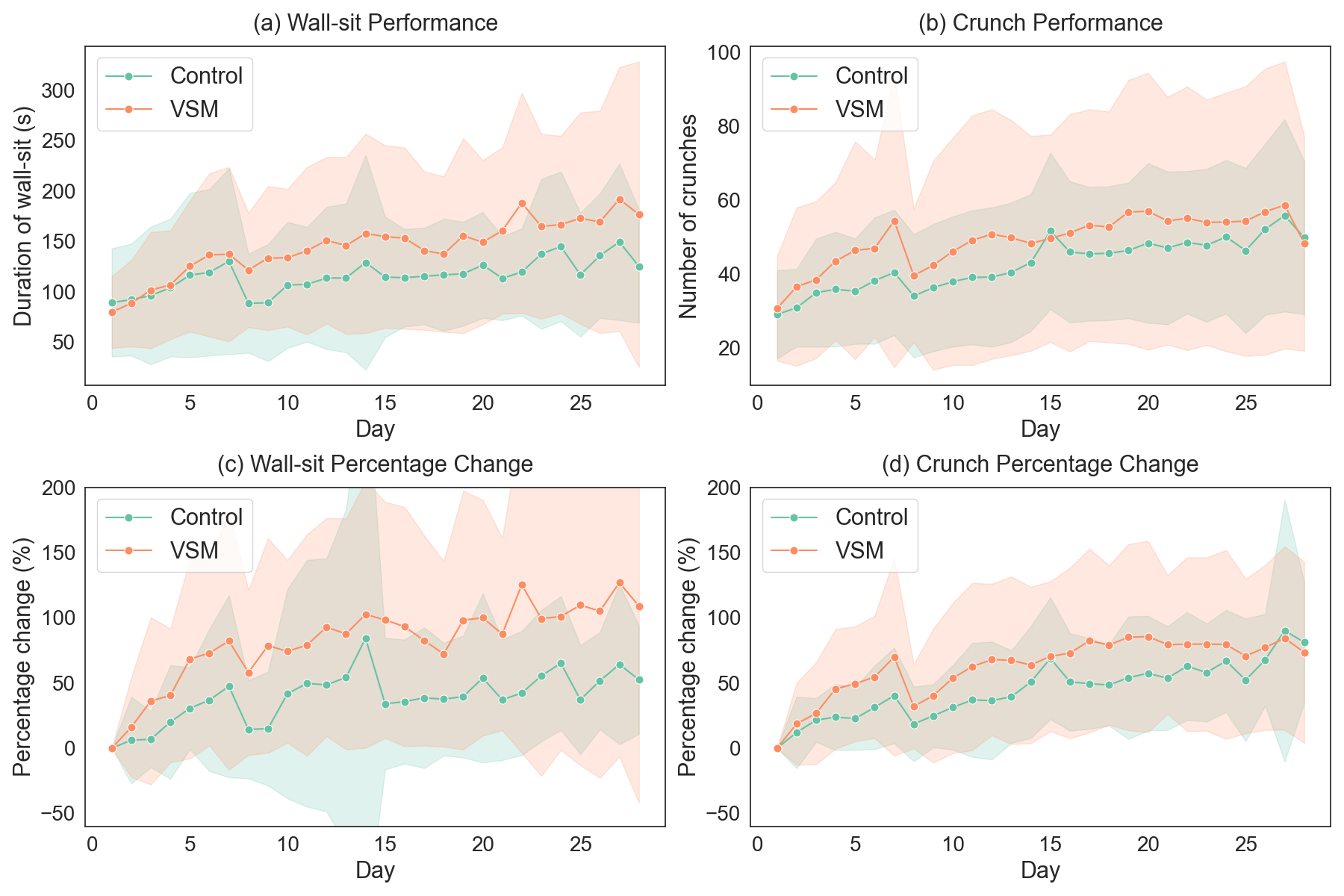} 
    \caption{Average objective performance and relative performance change for VSM and Control groups across four weeks.} 
    \label{fig:Study_2_performance}
    \Description{Four line graphs showing average performance and relative percentage change for VSM and Control groups across four weeks. (a) Wall-sit Performance: VSM consistently performs longer than Control. (b) Crunch Performance: VSM consistently completes more crunches than Control. (c) Wall-sit Percentage Change: VSM shows sharper relative improvement than Control. (d) Crunch Percentage Change: VSM maintains consistently greater relative improvement than Control.}
\end{figure*}

\begin{figure*}[t]
    \centering
    \includegraphics[width=0.8\textwidth]{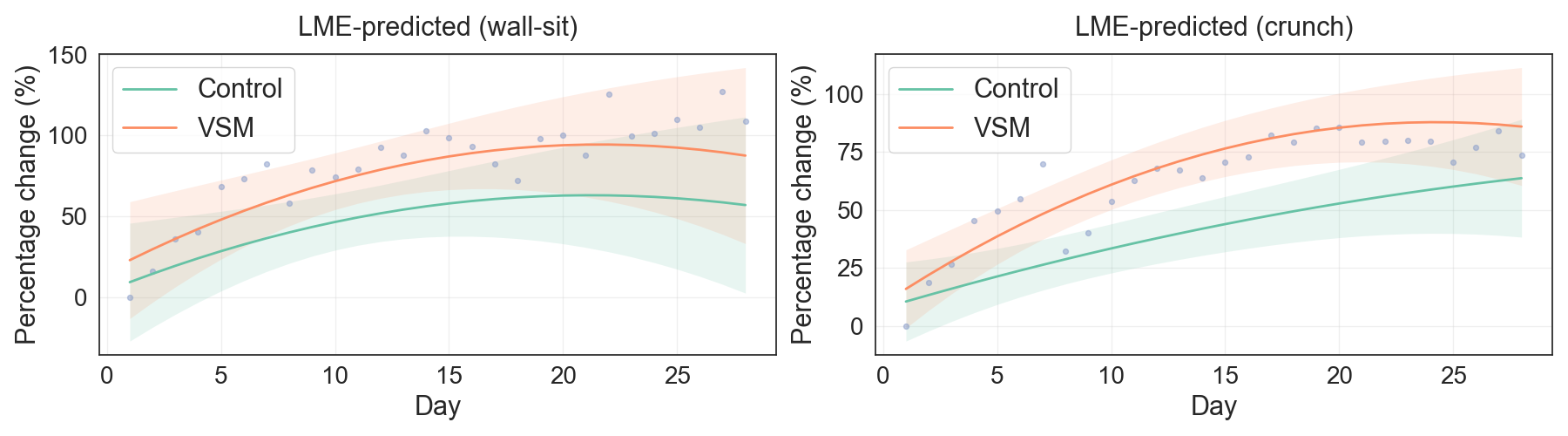} 
    \caption{Fitted performance curves over the one-month study based on the LME model.} 
    \label{fig:lme_predicted} 
    \Description{Two line graphs showing fitted performance trajectories over the one-month study based on the LME model. (a) LME-predicted Wall-sit: VSM shows a steeper upward trajectory, while Control rises more slowly. (b) LME-predicted Crunch: VSM shows stronger improvement, while Control follows a lower but steady upward trend.}
\end{figure*}

\subsection{RQ1: Does AI self-modeling (VSM) sustain performance over time?} \label{sec:study2_analysis_rq1}
We applied the same LME analysis methodology as described in Study~1 (Section~\ref{sec:exp1_analysis_methodology}) to assess performance in this one-month study, with the fitted curves shown in Figure~\ref{fig:lme_predicted}. Given the study length, we defined early and late stages as the average over 7 consecutive days (Days~1–7 and Days~22–28, respectively).

For wall-sit, results indicated a significant late advantage ($\Delta_L=31.34$, $SE=12.70$, $p=0.007$**), showing that the VSM group clearly outperformed the Control group in the final stage of the program. However, the sustainment contrast was not significant ($\Delta_L-\Delta_E=13.22$, $SE=15.18$, $p=0.192$), suggesting that the observed late-stage gap may not represent a statistically significant increase compared to the early-stage difference. For crunch, we again observed a robust late advantage ($\Delta_L=27.98$, $SE=7.94$, $p<0.001$**). The sustainment test showed a positive trend ($\Delta_L-\Delta_E=13.37$, $SE=9.48$, $p=0.079$), but did not reach conventional significance thresholds. Across both tasks, AI self-modeling (VSM) reliably produced superior performance at the late stage, confirming that its benefits did not vanish over time. Nevertheless, the evidence for sustaining or increasing relative gains is weaker: although trends point toward maintenance or slight growth of the advantage, the sustainment contrasts did not achieve strong statistical support. These findings suggest that VSM nudging secures lasting late-stage improvements, but further work is needed to verify whether such advantages consistently amplify with time.

\textbf{Key Findings:} In a four-week duration, VSM reliably leads to better performance at the late stage, whereas the sustainment of the relative gain is suggestive but not significant for both tasks.

\subsection{RQ2: Does AI self-modeling (VSM) sustain improvement rate over time?}
Study~1 showed that VSM did not sustain an accelerating rate of improvement; instead, performance gains tended to converge. To verify and characterize this convergence in a longer time frame, we combined LME analysis (Eq.~\ref{eq:analysis_rq2_beta5}) with subjective reports. 

\subsubsection{Stepwise Convergence Analysis.} \label{sec:study2_analysis_rq2_objective}
\begin{figure*}[t]
    \centering
    \includegraphics[width=\textwidth]{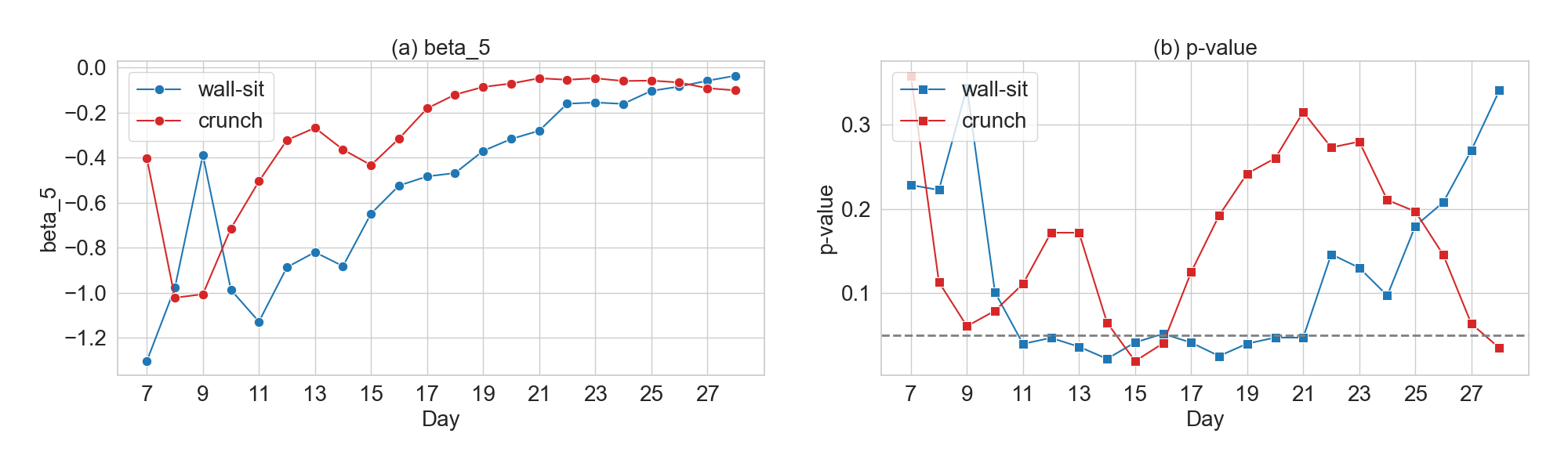} 
    \caption{Stepwise analysis of convergence: quadratic coefficients ($\beta_5$) and $p$-values across study days.} 
    \label{fig:step_fit}
    \Description{Two line graphs showing stepwise analysis of quadratic coefficients ($\beta_5$) and p-values for wall-sit and crunch tasks. (a) $\beta_5$ coefficients are consistently negative but trend toward zero over time for both tasks. (b) p-values fluctuate, with crunch showing greater variability and occasional peaks, while wall-sit remains mostly low, often below the 0.05 threshold.}
\end{figure*}
As illustrated in Figure~\ref{fig:Study_2_performance}, two distinct phases were suggested: an early rapid-growth phase for VSM, followed by convergence toward the Control group. To avoid obscuring such phase-specific dynamics and identify when the VSM growth rate began to decelerate, we adopted a stepwise, windowed strategy: we segment the series into progressively larger contiguous windows (7–28 days) and fit Eq.~\ref{eq:lme} within each window. The evolving pattern of $\beta_5$ (quadratic $Group \times t_c^2$ term) and its $p$-value is shown in Figure~\ref{fig:step_fit}, revealing the two-stage convergence pattern.


In the early to mid phase (Days~7--15), $\beta_5$ values for both exercises were negative and accompanied by gradually significant $p$-values ($p<0.050$ around Day~15 for crunch, Day~11-21 for wall-sit), indicating the existence of a decreasing improvement rate. This suggests that the strong initial acceleration of the VSM group's improvement began to weaken, reducing the improvement rate gap with Control. However, in the later phase (Days~21--28), $\beta_5$ continued to increase toward zero without significance. This fact implies that the VSM group's performance does not demonstrate a consistent parabolic trend; instead, their performance change possibly converged to the same trend as the Control group in the later phase. This shift suggests that while VSM sustained its early advantage throughout Study~2 (Section~\ref{sec:study2_analysis_rq1}), the marginal contribution of the intervention diminished over time.

\textbf{Key Findings:} The trajectory of the VSM group's performance is staged: an early rapid-growth burst when participants are first exposed to VSM nudging; a mid-phase convergence in which improvement rate decelerates (reflected by negative $\beta_5$); and a late phase (Days~21–28) where the VSM and Control groups' performances run roughly in parallel, implying attenuation of the VSM benefit.

\subsubsection{Subjective Motivation Analysis.}
\begin{figure*}[t]
    \centering
    \includegraphics[width=\textwidth]{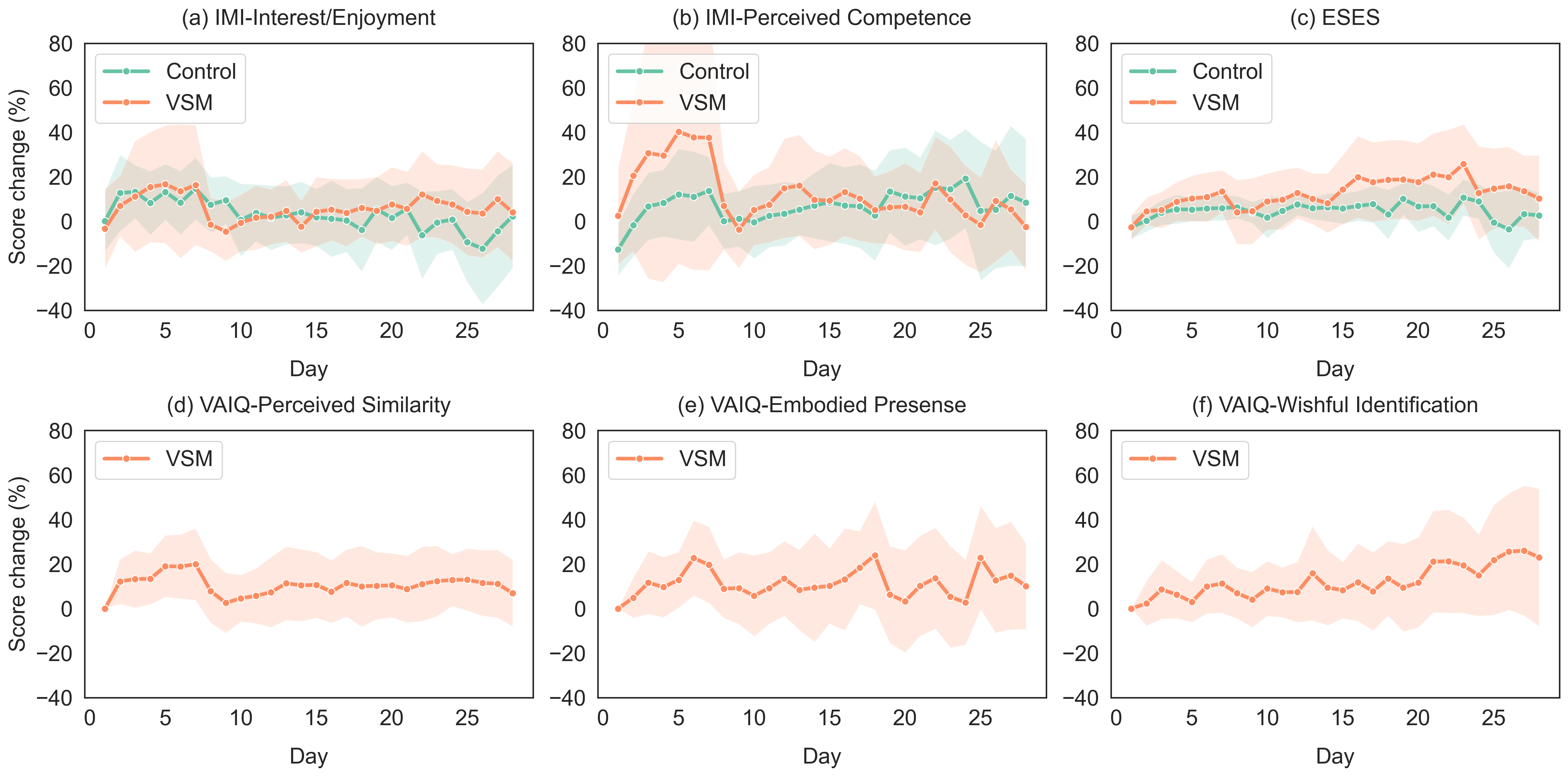} 
    \caption{Subjective measures of motivation and identification: changes in IMI, ESES, and VAIQ subscales over time.} 
    \label{fig:group_subjective}
    \Description{Six line graphs showing changes in subjective measures of motivation and identification (IMI, ESES, and VAIQ subscales) over the four-week study. (a) IMI–Interest/Enjoyment: both VSM and Control fluctuate around baseline, with VSM showing a slightly stronger upward trend early on. (b) IMI–Perceived Competence: VSM and Control follow similar trajectories, with no consistent separation. (c) ESES (Exercise Self-Efficacy Scale): both groups improve gradually, with VSM generally trending higher. (d) VAIQ–Perceived Similarity: VSM increases in the first week and then stabilizes. (e) VAIQ–Embodied Presence: VSM rises early and then declines gradually but remains above baseline. (f) VAIQ–Wishful Identification: VSM shows a steady upward trend throughout the study.}
\end{figure*}

Subjective measures provide additional insights into the mechanisms underlying the performance patterns observed in RQ1 and RQ2. As shown in Figure~\ref{fig:group_subjective}, participants’ psychological experiences followed a two-stage pattern: an initial novelty-driven boost, followed by attenuation and partial stabilization.

For the VSM group, both IMI-Interest/Enjoyment (Figure~\ref{fig:group_subjective}a) and IMI-Perceived Competence (Figure~\ref{fig:group_subjective}b) exhibited a sharp increase within the first week, peaking around Days 5–7 before declining toward control levels. This trend reflects the objective performance, suggesting that AI-generated videos initially stimulated excitement and competitiveness, but the motivational salience diminished with repeated daily exposure. Similarly, measures of identification experienced an early rise. VAIQ-Perceived Similarity (Figure~\ref{fig:group_subjective}d) and Embodied Presence (Figure~\ref{fig:group_subjective}e) increased markedly during the first week, then stabilized or fluctuated. These results indicate that novelty facilitated strong short-term self-identification, but its impact leveled off over time.

For the ESES (Figure~\ref{fig:group_subjective}c), the VSM group demonstrated an overall upward trend relative to Control, suggesting that although the direct nudging effect of video self-modeling attenuated over time, the intervention fostered a more durable sense of exercise self-efficacy. The slight decline observed in the final week may reflect attrition effects, as the number of participants had dropped to 15. Similarly, VAIQ-Wishful Identification (Figure~\ref{fig:group_subjective}f) showed a steady increase, indicating that participants’ pursuit of their ideal self and identification with the peer model became stronger over time. This sustained confidence and strengthened aspiration toward the ideal body explain why VSM participants maintained superior performance levels compared to the Control group even after novelty effects had diminished. In Section~\ref{sec:internalization}, we further examine these mechanisms through interviews and describe the process as an early catalyst followed by internalization.

\subsubsection{Conclusion.}
In this study, we extend the short-term findings of Study~1 and demonstrate that VSM can sustain higher performance levels across a four-week intervention. For RQ1, VSM participants maintained superior outcomes relative to Control, even after novelty effects diminished. RQ2, however, the improvement rate did not sustain; instead, performance gains followed a two-stage trajectory of early acceleration and later convergence. Subjective measures further reinforced this pattern, showing initial jumps of interest and identification that later stabilized into more durable self-efficacy and wishful identification.

\section{Design Implications}
\RI{Based on our post-study interviews and qualitative analysis of participants' reflections,} we distill three design implications for future behavior change technologies (BCTs). \RI{Interviews were fully transcribed and analyzed using Braun and Clarke’s six-phase reflexive thematic analysis~\cite{braun_2006}. Two researchers independently reviewed and coded the data before collaboratively refining a shared codebook. Detailed interview questions and our coding protocol are provided in Appendix~\ref{ap:code}. The resulting themes informed three implications:} (1) modality choices illustrated by the contrast between VSM and ASM; (2) personalized feedback informed by reflections on one's performance/subjective data; (3) participants’ direct design suggestions, which emphasize desired system features and interactive functions.
 
\subsection{VSM vs ASM: \RI{Embodiment} as Design Priority}
Study~1 revealed a significant contrast in effectiveness between VSM and ASM: VSM improved and sustained performance, while ASM failed to outperform the control condition. \RI{We posit that this difference stems from VSM's superior ability to facilitate embodiment, which is crucial for fitness tasks.} This contrast highlights three critical dimensions for future design: 

\RI{\textbf{Design for embodiment: enhancing motivation for physical tasks.} Embodiment is a powerful mechanism for motor learning and motivation. As established in embodied cognition and model-based learning research ~\cite{foglia2013embodied, shapiro_2019} and prior HCI work on visual guidance for movement ~\cite{semeraro2022visualizing, lin2021towards}, physical tasks rely on perceptual–motor mapping that is effectively supported through visual cues.} VSM’s effectiveness likely stems from its embodied visual presentation, providing a concrete, inimitable  “visual anchor” that helps participants map their own form, perceive gaps, and identify improvements. As P-V10 stated, \textit{``it feels like this is a body that I can achieve through training''}. In contrast, the auditory cues of ASM are disconnected from the user's physical experience, creating a cognitive and somatic gap. \RI{Bridging this gap may require moving beyond verbal cues to interactive audio systems, which have been shown to effectively foster embodiment~\cite{birchfield2008embodiment}.}

\textbf{Design for ``becoming'' instead of ``improving''}. \RI{Both ASM and VSM employed idealized self-models, while only VSM successfully fostered a sense of ``becoming'', achieving the internalization of the self-model as a plausible future identity that supports sustained motivation. VSM’s visual avatar naturally shifts users toward an aspirational identity, consistent with Possible Selves~\cite{markus_1986} and Identity-Based Motivation~\cite{oyserman_2015}. In contrast, ASM did not provide the necessary conditions for identity adoption. During the goal-setting phase, ASM participants frequently articulated process-oriented goals (e.g., ``be persistent,'' ``work harder'') rather than future-self outcomes, placing them in an ``improving” mindset rather than a ``becoming'' one. As P-A2 explained, auditory cues had a limited lasting impact: \textit{``Once you start exercising, you basically won’t think about it anymore.''} Similarly, P-A6 viewed the cues as transient support: \textit{``I prefer it gives me some encouragement words during the exercise.''} This suggests that the key design challenge for ASM is how to guide users toward a more embodied and identity-aligned experience, rather than merely offering momentary performance prompts.}


\textbf{Fidelity must surpass a threshold to create immersion.} 
Our study reveals the existence of a ``fidelity threshold'': while VSM's face-swapping was imperfect, its fidelity was sufficient for immersion. \RI{For example, although participants P-V5, P-V6, P-V7, and P-V13 noted that the \textit{``facial expressions were unnatural compared to daily life,''} they consistently reported still being able to recognize themselves; as P-V5 stated, \textit{``I can still recognize myself in the videos.''}} Conversely, ASM's voice-cloning, perceived as ``not similar'', caused participants to feel alienated, turning motivation into distraction. \RI{P-A4 described the audio as \textit{``slightly deliberate, a little unnatural... with interjections and long pauses,''} which disrupted immersion.} Thus, systems need only surpass this threshold to sustain immersion, rather than achieve perfection. This may reflect a perceptual asymmetry challenge in ASM: we hear ourselves through a mix of air- and bone-conduction that makes the ``internal'' voice sound warmer and lower than purely air-conducted playback~\cite{porschmann_2000}.


\subsection{Personal Data Reflection: Tailored feedback, Self-referencing, and Dynamic Intervention}

Participant's reflection on their personal data reveal that a one-size-fits-all intervention is insufficient due to individual differences. The design of BCTs should shift towards personalization: understand and adapt to the unique exercise pace, psychological states, and motivations of users.

\textbf{Provide tailored feedback according to user archetypes.} \label{sec:archetype}
Through interviews, we identified three typical user archetypes: (1) ``progressives'', (2) ``maintainers'', and (3) ``context responders''. 
\RI{Progressives such as P-C2, P-V5, and P-V8 described a consistent desire to ``do a little more each day.'' P-C2 shared, \textit{``it became a habit that I added one or two more each day... and near the end I felt happy.''} P-V5 echoed this pattern: \textit{``I always set a small goal, just a few more crunches than yesterday.''} Similarly, P-V8 described: \textit{``Every day I wanted to push a bit further—six minutes, then seven, then eight... I kept setting higher goals, and it felt good to accomplish them.''} Maintainers, by contrast, preferred stable, repeatable targets; P-V9 noted setting a ``generally similar goal for each day.'' Meanwhile, context responders showed highly variable patterns driven by external factors. P-C7, for instance, tied his strongest performance to feeling unusually energetic on a particular day: \textit{``I remember feeling very energetic the next day, so I did a lot.''} These archetypes suggest that systems should provide differentiated} goals and feedback, \RI{especially} for quantifiable exercises like crunches. Take a step further, inferring a user's strategic archetype from their exercise trends could achieve a smoother interaction experience. 

\textbf{Emphasize self-referencing instead of peer comparison.} AI self-modeling enhances motivation by providing a self-referencing role model instead of peer comparison. Self-confidence is closely associated with perceived success, and self-referencing can help build up a positive feedback loop. \textit{``I can easily achieve the improvement goal (set by myself), feeling more confident''} (P-V5), \textit{``at first I could barely hold (wall-sit), then I could stay much longer, I felt my improvement and my confidence went up too''} (P-V11). Conversely, peer comparison can bring anxiety and harm to self-confidence in the long run. P-V8 lowered his competence score because \textit{``I sometimes wondered if others were doing better than me... my competence score would be lower''}. Peer comparison may work in the short term, but it introduces risk in the long run.

\textbf{Dynamic intervention for shifting motivational states.} 
Participants point out two motivational states: (1) a positive feedback loop of ``effort-progress-satisfaction'', (2) a negative feedback loop of ``boredom-stagnant-burden'' when exercise becomes a task. P-V12 described his positive feedback loop as \textit{``a kind of push that made me exercise... it was more like a stepwise improvement... getting closer to the video gave me a real sense of achievement''}. Oppositely, P-V6 fell into a negative feedback loop after novelty decayed - \textit{``my interest definitely went down and I felt the effect wasn’t noticeable anymore...treating it like a task totally dampened my enthusiasm''.} This suggests the need for a dynamic intervention strategy that pulls users out of the negative feedback loop.  

\subsection{Participant Feedback: Interactivity, Diversity, and Adaptive Goals}
Specified suggestions from participants offer valuable insights to polish AI self-modeling into a user-centered design. Their feedback highlights the need for an interactive, adaptive, and motivating companion, beyond static demonstrations.

\textbf{From static demonstration to interactive guidance.} 
Participants reported a sense of distance and a mechanical feel from the VSM system. As P-V11 suggested, \textit{``If the AI video could provide personalized feedback on my form, it would increase my interest and attention when watching it''}, requiring the system to move beyond simple video playback. Future designs could integrate real-time pose analysis to provide immediate corrections, upgrading the experience from watching to being guided. Introducing a follow-along mode could also create a sense of virtual companionship.

\textbf{Combating habituation with variety and progression.} 
Content repetitiveness was a key factor in the decline of user interest over time. P-V12 explicitly stated that he \textit{``lost interest in the later stages due to the high repetitiveness of the videos''} and suggested \textit{``minor updates to the exercises every week''}. Beyond varying content, highlighting users’ improvement can re-energize engagement and motivate continued effort. P-V13 expressed a desire to \textit{``see a quantitative growth curve of my physical metrics''}. 

\textbf{Adapting goals to user progress.} An ideal AI self-model presents a near-future self that modestly outperforms the user’s current state. Conversely, a static, perfect AI self-model can lead to frustration due to the perceivable gap in performance. P-V10 noted that \textit{``I kind of doubled my crunch...but the standard in the video still felt out of reach''}. A more effective strategy is dynamic avatars that synchronize user progress with adaptive goals to reinforce a sense of improvement.

\section{Discussion}
\subsection{From Nudging to Internalized Motivation}\label{sec:internalization}

Our findings suggest that the motivational nudging process of VSM unfolds in three stages, beginning with an early catalyst effect, followed by novelty decrease, and eventually stabilizing through internalization of an ``ideal self''. The process reflects a broader understanding of human behavior change: instead of being passively nudged by environmental cues, people can transform external signals into enduring motivation~\cite{ryan_2000}, allowing interventions to persist beyond habituation.

While both VSM and ASM were designed as identity-based nudges, only VSM exhibited this three-stage nudging process. ASM showed limited early gains but did not transition toward internalized motivation.

\textbf{Early catalyst: clear visual goal and attainability.} 
At the beginning, VSM provided an actionable movement model and a tangible aspirational target, elevating the starting baseline relative to Control. Unlike normal fitness influencer content that is hard to achieve, the AI-generated videos struck a balance between role-modeling and attainability. By presenting exemplars that felt achievable, VSM enhanced self-efficacy and perceived control~\cite{gartzia_2021, morgenroth_2015}, thereby accelerating early performance gains.
In this stage, participants benefited from the clarity of a concrete visual target, which helped translate abstract goals into an actionable effort.

\textbf{Novelty decrease: fading stimulation and lack of feedback.}
As the exposure continued, the impact of VSM nudges diminished. The repetitive nudging made the intervention more predictable, and the stimulation was no longer as significant. As habituation theory explained, the repeated exposure to the same stimulus reduces its motivational salience over time~\cite{thompson_1966, rankin_2009}. Beyond fading novelty, the absence of feedback, critical for achieving both effectiveness and persistence in behavior change~\cite{diclemente_2001, fishbach_2012, belschak_2009}, made the participant reinterpret the goal as distant or discouraging. This explains why the rate of improvement slowed and trended toward convergence.

\textbf{Internalization of the ``ideal self'': a durable self-standard outlives novelty decay.} 
Even as attention to external stimulation declined, VSM continued to shape behavior by shifting from an external prompt to an internalized self-standard. Unlike traditional BCTs that depend on extrinsic cues such as reminders or rewards—and thus rarely sustain long-term impact unless personally meaningful~\cite{ryan_2000}—VSM fostered a transition from responding to visual stimulation to interpreting the AI self-model as a durable self-standard. This finding reinforces the idea of identity-based motivation theory that people are more likely to persist when goals align with their sense of ``who I am'' or ``who I want to become''~\cite{oyserman_2015}. Rather than merely fading as habituation theory predicts~\cite{thompson_1966, rankin_2009}, our findings suggest a more nuanced understanding: external intervention can be reframed into self-referential standards that sustain motivation.

\subsection{Generalization to Diverse Application Domains}
Our study demonstrates that VSM can effectively enhance motivation and performance, as evidenced by sustained improvements in wall-sit and crunch, showing its potential for supporting fitness behaviors. In contrast, ASM exhibited more variable effects: while it did not uniformly underperform, there were cases in which it interfered with performance. These findings suggest that the effectiveness of AI self-modeling is both modality-dependent and task-dependent, motivating a closer examination of how modality and task characteristics shape whether self-modeling supports or hinders sustained engagement.

\textbf{From wall-sit and crunch to diverse fitness tasks.}
Wall-sit and crunch represent only two examples of physical exercise, but the underlying principle of VSM is not limited to these tasks. Many fitness goals are inherently tied to visual representation, such as posture, body shape, or movement quality. VSM can present an attainable ``ideal self'' in the most intuitive way, suggesting the potential to support motivations for other daily exercise routines. However, in more specialized and demanding performance contexts, such as athletic training or rehabilitation, where improvements may be less visible or require technical precision, the advantage of VSM remains uncertain. 

\textbf{Modality and tasks across broader domains.}
AI self-modeling can extend to tasks beyond fitness to other domains requiring sustained effort with a meaningfully modeled ``ideal self''. For instance, in skill learning, learners may benefit from seeing and hearing themselves playing an instrument or speaking other languages. In mental health, a calmer or more confident future self could counter negative emotions. In lifestyle changes such as diet or study habits, self-modeling may strengthen identity-linked goals.

A key question for future application is how to match modality to domain. Our findings suggest that effectiveness depends on the representation of activities. Visually demonstrable tasks like fitness, sports, and diet are more likely to benefit from VSM, whereas ASM may have more potential in acoustic-related domains such as music, language learning, or emotional support. Exploring these applications can broaden the utility of AI self-modeling and clarify modality-specific boundaries.

\subsection{Limitation and Future Work}\label{sec:limitation}

\RIII{\textbf{Task selection and task–modality alignment.}} We selected wall-sit and crunch for their clear, quantifiable performance metrics, which strengthened the reliability of daily longitudinal measurement. However, both tasks rely heavily on posture and visual form. This task choice may have limited the robustness of our evaluation and may have interacted differently with the two implementations of self-modeling. As noted in Study~1, posture-dependent exercises offer stronger visual affordances than auditory ones, which may partly explain the weaker and more variable improvements seen with ASM.

\RIII{To fully understand whether audio-based self-modeling exhibits motivational patterns comparable to video-based self-modeling, future studies should examine task types that align more closely with auditory cues (e.g., endurance pacing, rhythm-based training, or internal-state tasks). Such tasks would allow a more balanced evaluation of modality–task fit and help determine whether the observed differences stem from the modality itself or from the characteristics of the selected exercises.}

\textbf{Ambiguous effects of ASM: support or hinder?} ASM underperformed the Control group in wall-sit and outperformed in crunch, without significance. As several participants noted, auditory cues quickly faded from attention once exercising began. Other participants also reported that the audio clip sounds funny to them. It is unclear whether ASM supports or hinders exercise tasks, and for what reason. Combined with the previous limitation, future work should investigate deeper into the properties and effects of ASM, as well as explore the reasons behind it.

\RII{\textbf{Participant motivation and monetary compensation:} Although financial compensation is common in long-term HCI studies, it is part of the reason that contributed to sustaining participant engagement. Several participants reported that the target exercises were not aligned with their personal fitness goals, suggesting that intrinsic motivation may not have been fully activated. Consequently, monetary incentives may have overshadowed the motivational contribution of self-modeling, especially in later stages of the intervention.}

\RII{This suggests a more realistic scenario. The next step of the research should evaluate self-modeling in voluntary product settings without monetary rewards. For instance, embedding VSM into community fitness challenges, goal-tracking apps, or wearable-based health programs where users already have an intrinsic interest. Long-term deployments in such naturalistic environments would allow a clearer understanding of self-modeling’s genuine motivational impact with fewer interventions of other factors.}

\RII{\textbf{Limitations in AI self-modeling technology.} The underlying technology for both VSM and ASM had practical imperfections. For VSM, some of the generated synthesized faces were occasionally unsettling, leading to a uncanny valley effect, while some were too unrealistic to be fully motivating. These reactions point to a well-recognized challenge in embodied AI design: self-representations that are too accurate can appear eerie, while those that are too stylized can weaken identification. Determining the optimal representational fidelity that is recognizable, motivational, and non-threatening remains an open design problem for video-based self-modeling.}

\RII{For ASM, fidelity issues arose from both technological and perceptual sources. Voice similarity varied across individuals, and also, the voice one hears internally differs from the voice captured externally due to bone conduction and resonance effects in human hearing. As a result, even accurately cloned voices sometimes felt ``not like me,'' reducing the sense of identification with the audio self-modeling. Addressing this gap may require voice transformation methods that approximate users’ self-perceived voice, or interactive calibration that allows users to adjust vocal timbre toward a more self-congruent representation.}

\RII{These technological and perceptual constraints likely influenced how strongly participants engaged with the self-modeling content. Future systems should explore balanced and customizable visual realism, as well as voice cloning calibrated to users’ self-perceived identity. Beyond fidelity improvements, expanding from static interventions to interactive and adaptive systems, such as multimodal self-modeling, context-aware feedback, or dynamically generated goals, may better sustain engagement over extended periods and support the transition from externally prompted nudges to internalized, self-directed motivation.}

\section{Conclusion}
This work conducted one of the first long-term empirical evaluations of AI self-modeling in the fitness domain. \RII{The first one-week study with 28 participants confirmed video self-modeling's (VSM) effectiveness, while audio self-modeling (ASM) failed to provide benefits for our setting, confirming the experiment setup for the next longitudinal study. Then, a 4-week study with 31 participants demonstrated that VSM can sustain higher performance levels but failed to maintain an accelerating improvement rate. By triangulating objective performance data with subjective measures and post-study interviews,} we show that the nudging effect of AI self-modeling can be diminishing yet internalized into lasting motivation, leading to design implications for future behavior change technologies.

\begin{acks}
This paper is supported by the National Key R\&D Program under Grant No. 2024YFB4505500 \& 2024YFB4505503, National Natural Science Foundation of China under Grant No. 62132010, 62472244, Qinghai University Research Ability Enhancement Project under Grant No. 2025KTSA05, CCF-Lenovo Blue Ocean Research Fund, Tsinghua University Initiative Scientific Research Program, and Undergraduate Education Innovation Grants, Tsinghua University.

\end{acks}

\bibliographystyle{ACM-Reference-Format}
\bibliography{nudging}

\appendix

\section{System Implementation}
\subsection{Pre-recorded Scripts for Voice Cloning}\label{ap:scripts}
Following prior work \cite{fang_2025_ESV}, we adapted the scripts to the local context. To ensure natural expression and emotional authenticity, participants read the corresponding version translated into their native language.

Read the following sentences with a happy tone:
\begin{itemize}
\item Hi there! Isn’t the weather nice today? The sunshine is so bright, and I feel completely energized!
\item I’ve got some great news! I received the job offer I’ve been dreaming of, and I’m so happy I can’t stop smiling!
\item Wow, this cake is just too delicious! You have to try it. It might be the best I’ve ever had!
\item Guess what? We’re going to the beach this weekend! I just can’t wait any longer!
\end{itemize}

Read the following sentences with a sad tone:
\begin{itemize}
\item What should I do? I can’t believe it… my pet ran away. I’ve checked everywhere but still couldn’t find it.
\item I’ve been feeling down all day today. I really wish things would get better soon.
\item This morning I got a call saying I didn’t pass the exam. I’ve felt awful the whole day.
\item Why every time I try so hard, the results are still disappointing? Sometimes it’s really difficult to stay optimistic.
\end{itemize}

\RIII{
\subsection{Prompt for Audio Content Generation}\label{ap:prompt}
Following prior work \cite{fang_2025_ESV}, we adapted the scripts to the local context. To ensure natural expression and emotional authenticity, participants read the corresponding version translated into their native language. The prompt template used in our study is provided below:}
\begin{quote}
\small
\ttfamily
SYSTEM\_PROMPT = (
\begin{list}{}{\leftmargin=1.5em \rightmargin=0em}
      \item Your task is to imagine yourself as the person with these trait personalities would say to themselves in the given scenario that would encourage them to keep up with the habit. You should try to embody the person when their habit has become their identity. Use the template of ``I am a xxx person''.
      \item You will also be given some settings to finetune the emotional affect of sentence: positivity, emotional expressivity. The default value is 0 and the range is -3 to +3.
      \item The personalities have more priority than the settings. 
      \item Requirements: You must express the attitudes and emotions saliently. You can add vocal bursts, natural vocal inflections and discourse markers. Never use markdowns or emojis. Use first-person.
      \item Keep the response short within four sentences.)
\end{list}

USER\_PROMPT = ( 
\begin{list}{}{\leftmargin=1.5em \rightmargin=0em}
        \item Scenario: {SCENARIO},
        \item Ideal self (5 words): {IDEAL\_SELF},
        \item Positivity: {POSITIVITY};
        \item Emotional expressivity: {EXPRESSIVITY})
\end{list}
\end{quote}

\section{User Questionnaires}
\subsection{Demographics}\label{ap:demographic}
\begin{itemize}
    \item Age
    \item Gender
    \item Weekly exercise frequency
    \item Main type of exercise
    \item Average duration per workout
    \item Habit of exercising at home
    \item Prior experience with AI-synthesized content (Question: "Have you ever watched or used AI-synthesized video or voice content?")
    \item Self-rated understanding of AI-synthesized content (Scale: –5 = unfamiliar, 5 = familiar)
\end{itemize}

\subsection{Daily Audio Generation Questionnaire}\label{ap:AGQ}
\begin{itemize}
    \item Please describe what kind of goal you would like to achieve in your exercise today.
    \item Please use five words to describe your ideal self.
    \item How positive would you like today’s generated audio to be? (Scale: –3 = negative, 3 = positive)
    \item How intense would you like the emotional expression in today’s generated audio to be? (Scale: –3 = low intensity, 3 = high intensity)
\end{itemize}

\subsection{Intrinsic Motivation Inventory (IMI)}\label{ap:IMI}
All questions were rated on a 7-point Likert scale, with items marked $\dagger$ included in the shortened daily questionnaire.

\textbf{Interest/Enjoyment}:
\begin{itemize}
    \item I enjoy this activity very much.$\dagger$
    \item This activity did not hold my attention at all.
    \item I would describe this activity as very interesting.$\dagger$
    \item I found this activity quite enjoyable.
    \item While I was doing this activity, I was thinking about how much I enjoyed it.
\end{itemize}

\textbf{Perceived Competence}:
\begin{itemize}
    \item I think I did pretty well at this activity.$\dagger$
    \item After doing this activity for a while, I felt quite competent.$\dagger$
    \item I am satisfied with my performance in this task.
    \item I was quite skilled at this activity.
    \item This is an activity that I am not very good at.
\end{itemize}

\subsection{Exercise Self-Efficacy Scale (ESES)}\label{ap:ESES}
All questions were rated on a 4-point Likert scale, with items marked $\dagger$ included in the shortened daily questionnaire.

\begin{itemize}
    \item If I try hard enough, I can overcome the barriers and challenges I encounter in physical activity and exercise.$\dagger$
    \item I am able to find ways and means to engage in physical activity and exercise.
    \item I can achieve the physical activity and exercise goals I set for myself.$\dagger$
    \item When I face barriers to physical activity or exercise, I can find multiple solutions to overcome them.
    \item I can engage in physical activity or exercise even when I feel tired.$\dagger$
    \item I can engage in physical activity or exercise even when I feel down.$\dagger$
    \item I can engage in physical activity or exercise even without the support of my family or friends.
    \item I can engage in physical activity or exercise even without the help of a therapist or a coach.$\dagger$
    \item I can motivate myself to restart physical activity or exercise even after taking a break for a while.$\dagger$
    \item I can engage in physical activity or exercise even if I don't have access to a gym, exercise facilities, or rehabilitation centers.
\end{itemize}

\subsection{Video/Audio Identification Questionnaire (VAIQ)}\label{ap:VAIQ}
All questions were rated on a 7-point Likert scale, with items marked $\dagger$ included in the shortened daily questionnaire.

\textbf{Perceived Similarity}:
\begin{itemize}
    \item (Audio) The AI-generated voice is like me in many ways.$\dagger$
    \item (Audio) The AI-generated voice resembles me.
    \item (Audio) I identify with the AI-generated voice.$\dagger$
    \item (Audio) I feel that this AI voice is very close to my own personality and temperament.
    \item (Video) The AI-generated avatar in the video is like me in many ways.$\dagger$
    \item (Video) The AI-generated avatar in the video resembles me.
    \item (Video) I identify with the AI-generated avatar.$\dagger$
    \item (Video) I feel that this AI avatar is very close to my own personality and temperament.
\end{itemize}

\textbf{Embodied Presence}:
\begin{itemize}
    \item (Video) When I see the video generated with the AI avatar, I feel like I am the person in the frame.
    \item (Video) Looking at the AI-generated me, I feel as if I have entered the character's body.
    \item (Video) While watching this video, I feel as if I have become one with the avatar.$\dagger$
    \item (Video) I feel that the body in the AI video is my own body.
\end{itemize}

\textbf{Wishful Identification}:
\begin{itemize}
    \item (Video) I wish I could be more like this AI-generated avatar.
    \item (Video) This AI avatar represents my ideal self.$\dagger$
    \item (Video) The AI-generated me is a "better me."
    \item (Video) This AI avatar has characteristics that I wish I had.
\end{itemize}

\section{Post-study Interview}\label{ap:code}
\subsection{Experience Rating Questions}
Participants were asked to rate the following items on a 7-point scale (Scale: 1 = lowest possible degree, 7 = highest possible degree).
\begin{itemize}
    \item How physically demanding was it to complete the daily exercise tasks throughout the entire study period?

    \item Did you feel the overall pace of the study protocol (including daily exercise, logging, and submission) was too compact or rushed?

    \item To what extent did you feel the study protocol aligned with your personal exercise goals and circumstances?

    \item Overall, how satisfied were you with your experience throughout the study period?
    \item Please answer the version of the following question that applies to your group:
    \begin{itemize}
        \item (Video) To what extent did the AI videos help you better understand and adhere to the exercises? 
        \item (Control) To what extent did the researchers' assistance help you complete the exercises? 
    \end{itemize}
    \item Did you find it difficult to adapt to the content and pace of the study?
    \item Was it difficult to complete the daily exercise and submission tasks during the study?
    \item What was your main motivation for participating in the study at the beginning? If you were to rate this motivation on a scale of 1-7, what score would you give? Did this motivation change over time?
    \item How significant a role did the monetary reward play for you? If there were no reward, would you still have persisted?
    \item Did you experience any periods of fatigue, numbness, or boredom? If so, how frequently did this occur (e.g., how many days a week)?
    \item If there were similar studies or long-term exercise programs in the future, how likely would you be to participate again?

\end{itemize}

\subsection{Open Questions}
\begin{itemize}
    \item What factors do you think influenced your motivation curve? Can you recall the reasons for each rise and fall? [\textit{Show the participant their objective and subjective charts}]

    \item Why do you think those rises and falls occurred?

    \item What impact, if any, has this experience had on forming a fitness habit or a new life rhythm for you?

    \item (For participants who dropped out) At what moment did you decide to stop? How were you feeling at that time?

    \item (Video) Do you usually watch fitness videos? If so, what types do you mainly watch? 

    \item (Video) What differences did you notice between the AI-generated videos and regular fitness videos? What aspects did you like or dislike? 

    \item (Video) During the study, were there any moments when you felt like quitting, similar to your past experiences with exercise? If so, what were the reasons, and do you feel the AI video had any motivational effect on you? 
\end{itemize}

\RII{\subsection{Qualitative Coding Procedure}
To analyze the semi-structured exit interviews from both Study~1 and Study~2, we followed Braun and Clarke’s six-phase reflexive thematic analysis approach~\cite{braun_2006}. Our goal was to understand participants’ subjective experiences with AI self-modeling, clarify the mechanisms underlying performance changes, and find design implications grounded in participants’ reflections.}

\RII{All interviews were audio-recorded, fully transcribed, and anony\-mized. Transcripts were then segmented into meaning units and imported into a qualitative coding environment.}

\RII{\textbf{Initial Coding}: two researchers independently conducted open coding, identifying segments segments that (a) had same or highly similar point raised by multiple participants, or (b) were insightful or informative for understanding AI self-modeling and its usage for long-term nudging. Codes remained close to participants’ language.}

\RI{\textbf{Theme Development}: Codes were iteratively clustered into broader themes. Because participants discussed both learning-related changes (e.g., posture awareness, technique refinement) and motivational dynamics (e.g., self-efficacy, identity cues, novelty), these mechanisms were intentionally separated. During this process, one recurring theme concerned how participants approached their daily goals. Some aimed for steady improvement, others for maintaining consistency, and others fluctuated with external constraints. These patterns were subsequently consolidated as the three ``user archetypes'' described in design implication (Section~\ref{sec:archetype}), emerging directly from repeated codes rather than being predefined categories.}

\RII{\textbf{Cross-Study Analysis}: the analyses of Study~1 and Study~2 followed the same high-level structure but were allowed to diverge when long-term patterns appeared. Study~1 predominantly reflected short-term responses such as modality–task alignment or initial fidelity impressions, whereas Study~2 introduced new themes related to longitudinal dynamics (e.g., early acceleration, stabilization, internalization).}

\RII{\textbf{Thematic Consolidation for Design Implications.}
Themes identified across studies informed the interpretation of behavioral change stages, clarified mechanisms behind performance changes, and grounded the design implications, such as the role of representational fidelity and the emergence of user archetypes. Representative quotes were selected to illustrate each theme and ensure transparency in how findings were derived.}

\RII{\section{Data Integrity and Attrition Details}\label{ap:attrition}
\subsection{Attrition Analysis: Tables and Survival Curves} 
This appendix details the statistical procedures used to validate the integrity of the dataset. We examined attrition from three perspectives: (1) Proportions: whether dropout rates differed by group or cohort; (2) Timing: whether dropout occurred earlier in specific groups; and (3) Characteristics: whether those who dropped out differed systematically from those who remained.}

\RII{\subsubsection{Dropout Contingency Tables and Fisher Exact Test}
Table~\ref{tab:dropout} summarizes the number of participants who completed or dropped out of the study across each $Cohort \times Group$ combination. ``Dropped'' denotes participants who did not finish all 28 study days.}

\RII{To test whether attrition rates differed systematically, we applied Fisher exact tests (appropriate given the small cell counts). Results showed no evidence of differential dropout: for $Group \times Dropout$, $OR = 1.02$, $p > 0.050$; for $Cohort \times Dropout$, $OR = 2.10$, $p > 0.050$, indicating that neither cohort membership nor intervention condition was associated with higher or lower dropout probability.}

\begin{table}[h]
\caption{\RII{Dropout counts by cohort and group.}}
\Description{Dropout counts by cohort and group. In Cohort 1, the Control group had 6 participants complete the study and 3 participants drop out. The Video group in Cohort 1 had 5 participants complete the study and 4 participants drop out. In Cohort 2, the Control group had 2 participants complete and 4 participants drop out. The Video group in Cohort 2 had 4 participants complete and 4 participants drop out.}
\label{tab:dropout}
\begin{tabular*}{\linewidth}{@{\extracolsep{\fill}} c c c c @{}}
\toprule
\textbf{Cohort} & \textbf{Group} & \textbf{Completed} & \textbf{Dropped} \\
\midrule
1 & Control & 6 & 3 \\
1 & Video & 5 & 4 \\
2 & Control & 2 & 4 \\
2 & Video & 4 & 4 \\
\bottomrule
\end{tabular*}
\end{table}

\begin{figure}[h]
    \centering
    \includegraphics[width=\columnwidth]{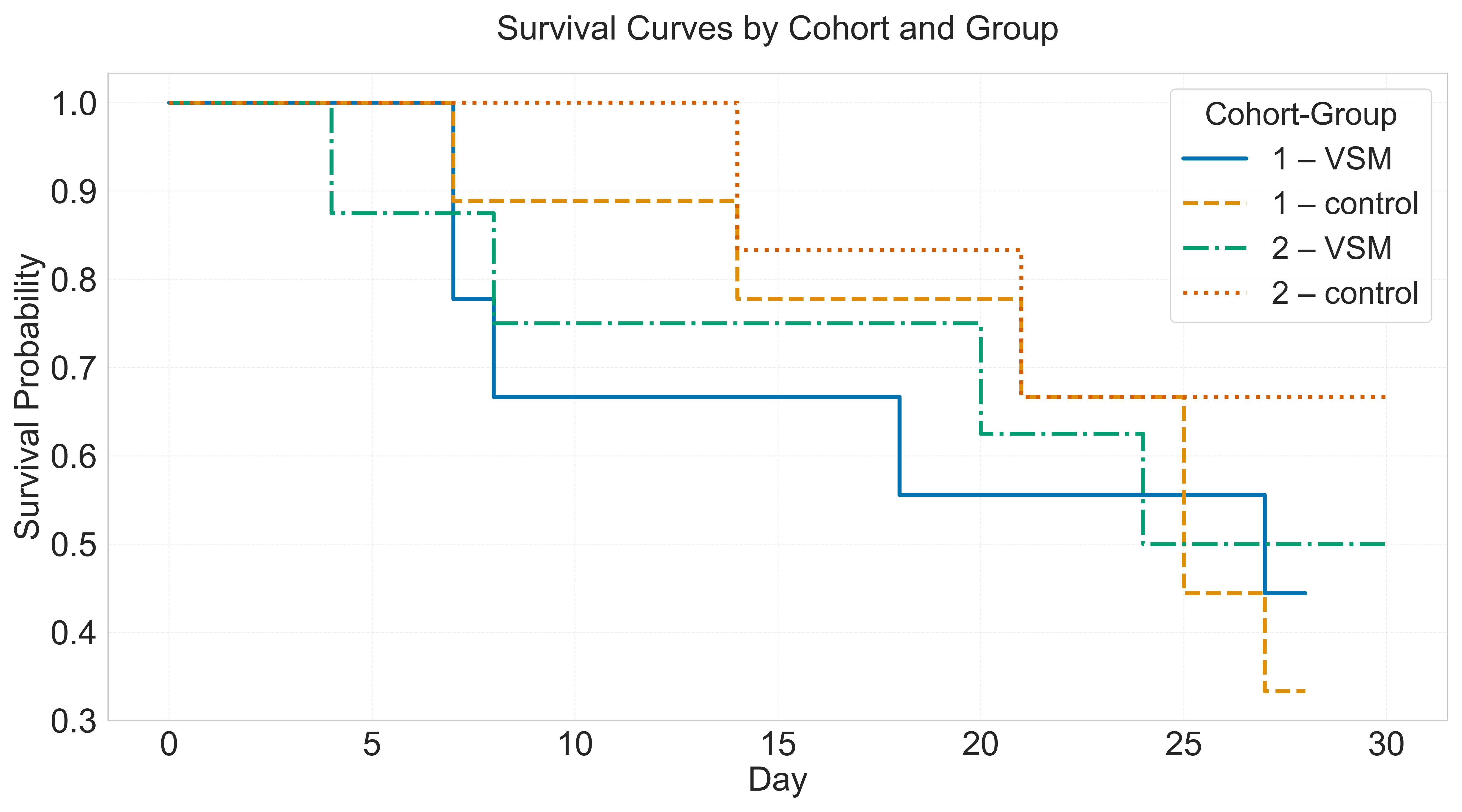} 
    \caption{\RII{Kaplan–Meier survival curves for Cohort~1 and Cohort~2 over the 28-day study period.}}
    \Description{Kaplan-Meier survival curves for Cohort 1 and Cohort 2 over the 28-day study period. The curves display the survival probability for four groups (Cohort 1-VSM, Cohort 1-control, Cohort 2-VSM, Cohort 2-control).}
    \label{fig:survival_curve}
\end{figure}

\RII{\subsubsection{Kaplan–Meier Survival Curve and Log-Rank Test}
We next examined whether the *timing* of attrition differed across cohorts using Kaplan–Meier survival analysis, where “survival” indicates remaining active in the study (i.e., not yet dropped out). Figure~\ref{fig:survival_curve} shows the survival functions for Cohort~1 and Cohort~2. The two curves track each other closely, suggesting no noticeable separation in dropout timing.}

\RII{We further used a log-rank test to statistically compared the survival distributions. For both the VSM group ($\chi^2 = 0.05$, $p = 0.820$) and the Control group ($\chi^2 = 1.07$, $p = 0.301$), the log-rank tests indicated no significant differences between Cohort 1 and Cohort 2. These results further support that attrition timing was not cohort-dependent.}

\RII{
\subsection{Survivor Bias Check}
To assess survivor bias, we performed independent t-tests comparing baseline metrics between Completers and Dropouts. The detailed statistical results are presented in Table~\ref{tab:survivor_bias}.}

\begin{table}[h]
\centering
\caption{\RII{Comparison of baseline characteristics between Completers and Dropouts within VSM and Control groups. (Mean $\pm$ SD)}}\label{tab:survivor_bias}
\Description{The table compares different metrics (IMI-Interest, IMI-Competence, ESES, Crunch, and Wall-Sit) between participants who completed the study and those who dropped out, separated into VSM and Control groups. For the VSM group, t-tests show no significant differences between completers and dropouts across any metrics. For the Control group, completers had significantly higher baseline scores in IMI-Interest/Enjoyment (t-value 4.76) and IMI-Perceived Competence (t-value 2.18) compared to dropouts. Physical metrics (Crunch, Wall-Sit) showed no significant differences in either group.}
\resizebox{\columnwidth}{!}{
\begin{tabular}{lcccccc}
\toprule
& \multicolumn{3}{c}{\textbf{VSM}}\\
\cmidrule(lr){2-4} \cmidrule(lr){5-7}
\textbf{Metric} & \textbf{Completed} & \textbf{Dropped} & \textbf{$t$-value}\\
\midrule
IMI-Interest/Enjoyment & 4.87 (0.89) & 4.71 (1.13) & 0.39\\
IMI-Perceived Competence & 4.90 (1.00) & 4.48 (1.29) & 0.92\\
ESES & 2.99 (0.38) & 3.18 (0.57) & -1.00\\
Crunch  & 44.10 (31.60) & 46.00 (16.60) & -0.18\\
Wall-Sit  & 117.90 (53.70) & 130.00 (68.40) & -0.49\\
\midrule
& \multicolumn{3}{c}{\textbf{Control}} \\
\cmidrule(lr){2-4} \cmidrule(lr){5-7}
\textbf{Metric} & \textbf{Completed} & \textbf{Dropped} & \textbf{$t$-value}\\
IMI-Interest/Enjoyment & 4.62 (0.68) & 3.40 (0.51) & 4.76$^{**}$ \\
IMI-Perceived Competence & 4.78 (0.77) & 4.09 (0.72) & 2.18$^{*}$ \\
ESES & 3.08 (0.37) & 3.02 (0.47) & 0.37 \\
Crunch  &  35.80 (14.40) & 36.30 (14.80) & -0.09 \\
Wall-Sit  & 108.10 (62.70) & 108.10 (72.10) & 0.00 \\
\bottomrule
\multicolumn{7}{l}{\footnotesize \textit{Note: Positive $t$-values indicate Completers > Dropouts.}}
\end{tabular}}
\end{table}

\end{document}